\documentclass[12pt]{article}
\usepackage{latexsym}
\usepackage{amsmath,amsbsy}

\newcommand{\dd}{\mbox{\rm d}}
\newcommand{\DD}{\mbox{\rm D}}

\newcommand{\oo}{\over}
\newcommand{\p}{\partial}
\newcommand{\vtheta}{\vartheta}
\newcommand{\vphi}{\varphi}
\newcommand{\ci}{\cite}

\newcommand{\be}{\begin{equation}}
\newcommand{\ee}{\end{equation}}
\newcommand{\lbl}{\label}
\newcommand{\bi}{\bibitem}

\newcommand{\vs}{\vspace}
\newcommand{\hs}{\hspace}

\newcommand{\pket}{|\mbox{${1\oo 2}{1\oo 2}$} \rangle}
\newcommand{\mket}{|\mbox{${1\oo 2},-{1\oo 2}$} \rangle}
\newcommand{\pbra}{\langle \mbox{${1\oo 2}{1\oo 2}$}|}
\newcommand{\mbra}{\langle \mbox{${1\oo 2},-{1\oo 2}$}|}

\def\bear{\begin{eqnarray}}
\def\ear{\end{eqnarray}}

\begin{document}
\baselineskip .7cm

\thispagestyle{empty}

\ 

\vs{1.5cm}

\begin{center}

{\bf \LARGE Rigid Particle and its Spin Revisited}

\vs{6mm}

Matej Pav\v si\v c\footnote{Email: Matej.Pavsic@ijs.si}

Jo\v zef Stefan Institute, Jamova 39, SI-1000 Ljubljana, Slovenia

\vs{1.5cm}

ABSTRACT
\end{center}

The arguments by Pandres that the double valued spherical harmonics
provide a basis for the irreducible spinor representation of the
three dimensional rotation group are further developed and justified.
The usual arguments against the inadmissibility of such functions,
concerning hermiticity, orthogonality, behavior under rotations, etc., are 
all shown to be related to the unsuitable choice of functions representing the
states with opposite projections of angular momentum. By a correct choice
of functions and definition of inner product those difficulties do not occur.
And yet the orbital angular momentum in the
ordinary configuration space can have integer eigenvalues only, for the reason
which have roots in the nature of quantum mechanics in such space.
The situation is different in the velocity space of the rigid particle,
whose action contains a term with the extrinsic curvature.

\newpage

\section{Introduction}

The theory of point particle whose
action contains not only the length, but also the extrinsic curvature of
the worldline has attracted much attention \ci{Rigid}--\ci{PavsicRigidConsist}. 
Such particle, commonly called ``rigid particle'', is a
particular case of rigid membranes of any dimension (called `` branes'').
The rigid particle behaves in all respects as a particle with spin. The spin
occurs because, even if free, the particle traces a worldline which
deviates from a straight line. In particular, it can be a helical
worldline \ci{PavsicRigid1}. 
In the absence of an external field, the constants of motion are
the linear momentum $p_\mu$ and the total angular momentum
$J_{\mu \nu}$ which is the sum of the orbital
angular momentum $L_{\mu \nu}$ and the spin $S_{\mu \nu}$. In the presence
of a gravitational field, the equation of motion for rigid particle was
shown \ci{PavsicRigid1} to be just the
Papapetrou equation \ci{Papapetrou}. The algebra of the (classical) Poisson
brackets and the (quantum) commutators resembles that of a spinning
particle and it was concluded that the rigid
particle leads to the Dirac equation \ci{PavsicRigid2}. In refs. 
\ci{Plyushchay,LindstromAccel} a counter argument occurred, namely that
the spin of the rigid particle is formally like
the orbital momentum, with the only difference that it acts not in the ordinary
configuration space, but in the space of velocities. Since orbital momentum
is well known to posses integer values only, it was concluded that the
rigid particle cannot have half-integer spin values. In the present paper
we will challenge that conclusion. 

A theoretical justification of why orbital angular momentum is allowed
to have integer values only, and not half-integer,
had turned out to be not so straightforward, and the arguments had 
changed during the course of investigation. Initially \ci{Schrodinger}
it was taken for granted that the wave function had
to be single valued. Then it was realized \ci{Pauli1}
that only experimental results needed
to be unique, but the wave function itself did not need to be single valued.
So Pauli \ci{Pauli2} found another argument, namely that the appropriate set
of basis functions has to provide a representation of the rotation group.
He argued that the spherical functions $Y_{lm}$ with half-integer $l$ fail
to provide such representation.

Amongst many subsequent papers 
\ci{Winter},\ci{Sanikov}--\ci{Whippman} on the subject
there are those by Pandres \ci{Pandres,Pandres1} who demonstrated
that the above
assertion by Pauli was not correct. Pandres conclusion was that the
functions $Y_{lm}$ with half-integer $l$ do provide the basis for an
irreducible representation of the rotation group. Pandres explicitly
stressed that he had no quarrel with Pauli's conclusion concerning
the inadmissibility  of multivalued quantum mechanical wave functions
in descriptions of the ordinary orbital angular momentum, although he 
took issue with the argument through which Pauli had reached that 
conclusion. In the following I am going to clarify and further develop
Pandres' arguments. In particular, I will show that although the usual
orbital angular momentum in coordinate space indeed cannot
have half-integer values,
the situation is different in the {\it velocity space} of the rigid particle.
In the velocity space the functions $Y_{lm}$ with half integer $l$ and $m$
are acceptable not only
because they do provide a basis for representation of the rotation group,
but also because the dynamics of the rigid particle, its equations of
motion and constants of motion, are different from those of a usual
quantum mechanical particle. So the linear momentum $\pi_\mu$
in velocity space is not a constant of motion and the eigenfunctions of the
operator ${\hat \pi}_\mu$ are not solutions of the wave equation
for the quantized rigid particle. Since it has turned out \ci{LindstromAccel}
much more convenient
to formulate the theory not in the velocity, but in the acceleration space,
I will explore the 'orbital angular momentum' operator in the latter space,
and show that its eigenvalues can be half-integers.

\section{The Schr\" odinger basis for spinor representation of the
three-dimensional rotation group}

Amongst many papers \ci{Pauli1}--\ci{Winter},\ci{Sanikov}--\ci{Whippman}
on angular momentum and its representation
the paper by Pandres \ci{Pandres}
---with the above title--- is distinct in claiming that
the rotation  group can be represented by means of double valued spherical
harmonics. I will re-examine his arguments and confirm that Pandres's
understanding of the problem was deeper from that of other researchers.
Half-integer spin is special ---in comparison with the integer spin--- 
in several respects, the most notorious being its property that a $2 \pi$
rotation does not bring the system in its original state:
the additional $2 \pi$
rotation is necessary if one wishes to arrive at the initial situation.
A spin $1\oo 2$ system has an orientation--entanglement with its
environment. This has consequences if one tries to describe the system
by employing the Schr\" odinger representation. One immediately finds out
that this cannot be done in the same way as in the case of a system with an
integer value of angular momentum. The spherical harmonics with
half-integer values do provide a basis for the irreducible spinor
representation of 
the three-dimensional rotation group, provided that one imposes
certain ``amendments" to what is meant by ``forming a representation''.
Such
 amendments should not be considered as unusual for spinors---which
are themselves unusual objects in comparison with the more ``usual"
objects---and are in close relation to orientation--entanglement
of a spinor object with its environment, which is illustrated in the well-known
example of a classical object attached to its surroundings by elastic
threads. Evidently, as stated by Misner et al.\,\ci{Wheeler},
in the case of spinors
there is something about the geometry of orientation that is not fully taken
into account in the usual concept of orientation.

Our first ammendment is related to the fact that under rotations
the spinorial wave functions do not transform as scalar functions.
The failure of half-integer spherical harmonics to behave as scalars
has been taken as one of the crucial reasons
to reject them in the description
of {\it orbital angular momentum}, which was indeed reasonable.
But if we want to use them in order to represent {\it spinors},
then obviously they should not
behave as scalars under rotations.

Our second ammendment is related
to the fact that half-integer spherical harmonics do not form a
fully irreducible representation of the rotation group. The total space
splits into an infinite dimensional and a finite dimensional subspace.
The latter subspace, denoted ${\cal S}_l$, is {\it not invariant};
if we start from a state, initially in ${\cal S}_l$, after a rotation
we obtain a state with components not only in ${\cal S}_l$, but also
outside ${\cal S}_l$. But it turns out that
the {\it projection} of the state onto ${\cal S}_l$ transforms just as
a spinor, its norm being preserved. The occurrence of the
states outside ${\cal S}_l$ has no influence on the behavior
of the projected state.
Therefore we can say that the basis states of ${\cal S}_l$ do form
a representation of the rotation group in such a generalized, ``ammendedd",
sense. This makes sense, because the states outside ${\cal S}_l$ are
unphysical, due to the fact that with respect to the latter states
the expectation value of the nonnegative definite operator
$L_x^2 + L_y^2$ is negative. So we have to disregard those ``ghost-like"
states, and we do this by performing the projection of a generic state onto
the physical space ${\cal S}_l$. We also show that an alternative way
of eliminating the unphysical states from the game is in adopting
a suitably renormalized inner product with a consequence that norms
of the unphysical states are zero.

\subsection{Choice of functions}

Let $L_i$ be a set of Schr\" odinger-type operators
\bear
    &&L_x = i ({\rm cot} \, \vartheta \, {\rm cos} \, \varphi \,
    {{\p}\oo {\p \varphi}}
   + {\rm sin} \, \varphi  \, {\p \oo {\p \vartheta}}) \nonumber \\
   && L_y = i ({\rm cot} \, \vartheta \, {\rm sin} \, \varphi \, 
   {{\p}\oo {\p \varphi}}
    - {\rm cos} \, \varphi  \, {\p \oo {\p \vartheta}}) \nonumber \\
  &&L_z = - i {\p \oo {\p \varphi}}
\lbl{1}
\ear
where $\vartheta$ and $\varphi$ are the usual polar coordinates.

Let us consider the functions $Y_{lm} (\vartheta, \varphi)$ which satisfy
the equation
     \be
              \boldsymbol{L}^2 Y_{lm} = l (l +1) Y_{lm}
\lbl{2}
\ee
\be
          L_z Y_{lm} = m Y_{lm}
\lbl{3}
\ee
where
 \be
           \boldsymbol{L}^2 \equiv L_x^2 +L_y^2 + L_z^2 = - {1\oo {{\rm sin}^2 \vartheta}}
      {{{\p}^2}\oo {\p \varphi^2}} - {1\oo {{\rm sin} \, 
      \vartheta}} {\p \oo {\p \vartheta}} \left ( {\rm sin} \, \vartheta \, 
    {\p \oo {\p \vartheta}} \right )
\lbl{2a}
\ee
For integer values of $l$ the $Y_{lm}$ are the familiar single-valued spherical
harmonics, whilst for half-integer values of $l$ the $Y_{lm}$ are double
valued functions.

In general, for any integer or half-integer values of $l$ the functions that satisfy
eqs.(\ref{2}),(\ref{3}) are given by
\be
   Y_{lm} = {1\oo \sqrt{2 \pi}} \, e^{im \varphi} (-1)^l \sqrt{{{2 l + 1}\oo
   2}} \sqrt{{{\Pi(l+m)}\oo {\Pi (l-m)}}} {1\oo {2^l \Pi (l)}}
   {1\oo {{\rm sin}^m \vartheta}} {\dd^{l-m} \oo 
   {(\dd \, {\rm cos} \, \vartheta)^{l-m}}} {\rm sin}^{2l} \vartheta
\lbl{4}
\ee  
\be
    m =  \left\{ \begin{array}{ll}
                l, l-1,..., -l & \mbox{if $l$ integer} \\
                l,l-1,...,-l, -l-1, -l-2,... & \mbox{if $l$ halh-integer}
                \end{array}
          \right.
\lbl{4a}
\ee
where $\Pi(l) \equiv \Gamma(l+1)$ is a generalization of $l!$ to non integer
values of $l$.

Besides (\ref{4}) there is another set of functions which solves the
system (\ref{2}),\,(\ref{3}):
\be
Z_{lm} = {1\oo \sqrt{2 \pi}} \, e^{im \varphi} (-1)^{(l+m)} \sqrt{{{2 l + 1}\oo
2}} \sqrt{{{\Pi(l-m)}\oo {\Pi (l+m)}}} {1\oo {2^l \Pi (l)}}
{1\oo {{\rm sin}^{-m} \vartheta}} {\dd^{l+m} \oo {(\dd \, {\rm cos} \, 
\vartheta)^{l+m}}} {\rm sin}^{2l} \vartheta
\lbl{5}
\ee
\be
    m = \left\{ \begin{array}{ll}
                -l, -l+1,..., l & \mbox{if $l$ integer}\\
                -l,l+1,...,l, l+1, l+2,... & \mbox{if $l$ halh-integer}
                \end{array}
        \right.
\lbl{5a}
\ee
The function $Z_{lm}$ coincide with $Y_{l,m}$ for 
integer values of $l$ only.
In the case of half-integer $l$-values, they are different.

If we define the raising and lowering operators as usually
\be
      L_x + i L_y \equiv L_+ = {\rm e}^{i \varphi} \left ( {\p \oo
      {\p \vartheta}} + i \, {\rm cot} \, \vartheta \, 
      {\p \oo {\p \varphi}} \right )
\lbl{6}
\ee
\be
      L_x - i L_y \equiv L_- = {\rm e}^{-i \varphi} \left ( -{\p \oo
      {\p \vartheta}} + i \, {\rm cot} \, \vartheta \,
      {\p \oo {\p \varphi}} \right )
\lbl{7}
\ee      
we find \ci{Pandres} for {\it half-integer values} of $l$
\be
    L_+ Y_{lm} \propto \left\{ \begin{array}{ll}
                               Y_{l,m+1} \; , & m = l-1,l-2,...,-l; \; -l-2,-l-3,... \\
                               0 \; , & m = l, \;  m = -l-1
                               \end{array}
                         \right.
\lbl{8}
\ee
\be
     L_- Y_{lm} \propto Y_{l,m-1} \; , \quad m = l,l-1,...,-l,-l-1,...
\lbl{9}
\ee
\be
     L_+ Z_{lm} \propto Z_{l,m+1} \; , \quad m = -l,-l+1,...,l,l+1,...
\lbl{10}
\ee
\be
    L_- Z_{lm} \propto \left\{ \begin{array}{ll}
                               Z_{l,m-1} \; , & m = -l+1,-l+2,...,l\,;\;                                l+2,l+3,... \\
                               0 \; , & m = -l, \;  m = l+1
                               \end{array}
                         \right.
\lbl{11}
\ee

Now let ${\cal S}_l$ be a function space which is spanned by the basis
functions $Y_{lm}$ for a given value of $l$ and for $m=-l,...,l$. Further,
let ${\cal O}_l$ be a space spanned by $Y_{lm}$ for a given value of $l$ and
for $m= -l-1,-l-2,...$.
Analogous we have for functions $Z_{lm}$.

From the relations (\ref{8})--(\ref{11}) we see that although the repeated
application
 of $L_-$ to $Y_{l,-l}$ does not give zero, but gives 
$Y_{l,-l-1},Y_{l,-l-2},...$,
i.e., brings us out of ${\cal S}_l$ into ${\cal O}_l$, the reverse is not
true. If $L_+$ is applied to $Y_{l,-l-1} \in {\cal O}_l$ the result is zero.
This comes directly from the identity
\be
    L_\pm L_\mp Y_{lm} = (l\pm m)(l\mp m + 1) Y_{lm} 
\lbl{a1}
\ee
from which we find
\be
      L_+ L_- Y_{l,-l} = 0
\lbl{a2}
\ee
Since $L_- Y_{l,-l} \propto Y_{l,-l-1}$ we have ffrom eq.(\ref{a2}) that
\be
L_+ Y_{l,-l-1} = 0
\lbl{a3}
\ee

Analogously, from $L_- L_+ Z_{l,l}=0$ we obtain the relation 
\be
    L_- Z_{l,l+1} = 0
\lbl{a4}
\ee

In particular we have,
\bear
    &&L_- Y_{{1\oo 2},-{1\oo 2}} \propto Y_{{1\oo 2},-{3\oo 2}} \lbl{12}\\  
    &&L_+ Y_{{1\oo 2},-{3\oo 2}} = 0 \lbl{13}
\ear
This can be verified by direct 
calculations using the differential operators (\ref{6}),(\ref{7}) 
and functions  (\ref{4}),(\ref{5}). For instance, taking
\be
    Y_{{1\oo 2},{-{3\oo 2}}} = L_- Y_{{1\oo 2},-{1\oo 2}} =
    - {i \oo \pi} \, {\rm sin}^{-3/2} \, \vartheta \, 
    {\rm e}^{-i {{3 \varphi}\oo 2}}
\lbl{13a}
\ee
we find
\be
    L_+ Y_{{1\oo 2},{-{3\oo 2}}} = {\rm e}^{i \varphi} \left ( {\p \oo
      {\p \vartheta}} + i \, {\rm cot} \, \vartheta \, 
      {\p \oo {\p \varphi}} \right )
      \left (- {i \oo \pi} \right ) \, {\rm sin}^{-3/2} \, \vartheta \, 
    {\rm e}^{-i {{3 \varphi}\oo 2}} = 0
\lbl{13b}
\ee  
Similar is true for the functions $Z_{lm}$, with the role of $L_+$ and
$L_-$ interchanged.

The spherical harmonics for half-integer $l$-values differ from those for
integer $l$-values in the properties such as those expressed in
eqs.\,(\ref{8}),(\ref{9}), i.e.,
\be
    L_- Y_{l,-l} \neq 0 \; , {\rm and } \quad L_+ Z_{l,l} \neq 0
\lbl{13c}
\ee
which imply the existence of functions outside the space
${\cal S}_l$. Those functions have negative eigenvalues of
the operator $\boldsymbol{L}^2 - L_z^2 = L_x^2 + L_y^2$, and therefore,
according to the ordinary criteria cannot be considered as representing physical
states\footnote{Analogously, negative norm states in gauge theories are
unphysical, and yet they can be consistently employed in the formulation
of the theories.}. Physical states are obtained by projecting an
arbitrary state into the subspace ${\cal S}_l$.

In the following I am going to show that using the functions 
(\ref{4}),(\ref{5}) with the properties (\ref{8})--(\ref{a4}) the usual 
arguments against such  functions as representing states with 
half-integer angular momentum do not hold. The confusion
has been spread into several directions. Besides Pauli's very sound 
argument concerning the behavior of the system under rotations 
there are other claims such as
:
\begin{itemize}
  
  \item functions $Y_{lm}$ and $Y_{l'm}$ are not orthogonal,
 
  \item they may have infinite norms,

   \item $L_i$ are not Hermitian,
 
  \item other problems \ci{Winter}.
  
\end{itemize}
I will now discuss those claims.

{\it Orthogonality -} One finds that not only the functions $Y_{lm}$
and $Y_{lm'}$, but also $Y_{lm}$
and $Y_{l'm}$, belonging to the set (\ref{4}),(\ref{5}) are orthogonal.
This is not the case for the set of functions used by Merzbacher and Van Winter
who start formally from
the same expression (\ref{4}), but restrict
the range of allowed $l,m$ values in a way different from (\ref{4a}), so that
for positive $m$-values they take functions (\ref{4}), whilst
for negative $m$-values they take functions (\ref{5})\footnote{
Such unsuitable set of functions has been recently used by Hunter et al.\,\ci{Hunter},
who otherwise correctly argued  that half integer spherical harmonics
can represent spin.}.

{\it Infinite norms} - However, a problem remains even
with our choice
of functions. Certain states of ${\cal S}_l$, namely those with
$m=-{3\oo 2},-{5\oo 2},...,-l$, have infinite norms, i.e.,
$\langle lm|lm \rangle$ is infinite. But, as stated by Pandres,
it is a well known fact that the inner product can be redefined so to
obtain finite norms, normalized to unity.

Let us consider the quantities
\be
     G_{m m''} (\epsilon) = \int_{\Omega-\epsilon} \dd \Omega \,
     Y_{lm}^* \, Y_{lm''}
\lbl{B1}
\ee
where we have performed a cut off in the integration domain. Instead of
integrating over the domain
\be
     \Omega = \lbrace (\varphi, \vartheta)| \varphi \in [0,2 \pi],~
     \vartheta \in [0,\pi] \rbrace
\lbl{B2}
\ee
we integrate over a truncated domain
\be
     \Omega -\epsilon = \lbrace (\varphi, \vartheta)| \varphi \in [0,2 \pi],~
     \vartheta \in [\epsilon,\pi -\epsilon] \rbrace
\lbl{B3}
\ee

The quantities $G_{m m''} (\epsilon)$ are zero, if $m\neq m''$, and
different from zero and finite, if $m=m''$. This is so also if
$m=-{3\oo 2},-{5\oo 2},...,-l$.

Now let $G^{m'' m} (\epsilon)$ be the inverse matrix to $G_{m m''}$.
So we have
\be
    \int_{\Omega-\epsilon} \dd \Omega \, Y_{lm}^* \, Y_{lm''}
G^{m'' m'} (\epsilon) = G_{m m''} (\epsilon) G^{m'' m'} (\epsilon)
= {\delta_m}^{m'}
\lbl{B4}
\ee

\paragraph{Definition I of inner product -}
The latter relation is valid for any value of $\epsilon$, whatever small.
Using eq.\,(\ref{B4}), we define the inner product between two functions
according to:
\be
   (Y_{lm}, Y_{lm'}) = \lim_{\epsilon \rightarrow 0} 
   \int_{\Omega-\epsilon} \dd \Omega \, Y_{lm}^* \, Y_{lm''}\,
G^{m'' m'} (\epsilon) = {\delta_m}^{m'}
\lbl{B5}
\ee
If $m=m'$, this can be written as 
\be
    (Y_{lm},Y_{lm}) = \lim_{\epsilon \to 0} 
    \frac{(Y_{lm},Y_{lm})_\epsilon}{N_{lm}(\epsilon)}
   \quad ,  \qquad N_{lm}(\epsilon) = {(Y_{lm},Y_{lm})_\epsilon}
\lbl{B6}
\ee   
For values of $m$ and $m'$ other than $\lbrace -{3\oo 2},-{5\oo 2},...,-l
\rbrace$, we have $\lim_{\epsilon \to 0} G_{m m''} (\epsilon) =
\delta_{m m''}$ and $\lim_{\epsilon \to 0} G^{m'' m'} (\epsilon)
 = \delta^{m'' m'}$,
so that in this particular case the inner product coincides with
the usual inner product.

{\it Hermiticity -} By using the set of functions (\ref{4}),(\ref{5})
and  the relations
(\ref{8})--(\ref{11}) one finds that the operators $L_i$,
$\boldsymbol{L}^2$ are Hermitian with respect to ${\cal S}_l$.
This is not the case  if one uses a different
set of functions---as Merzbacher \ci{Merzbacher} and Van Winter \ci{Winter}
did---such that, e.g.,  ${\cal S}_l$ for $l=\frac{1}{2}$ consists of
\bear
      && Y_{{1\oo 2} {1\oo 2}} \propto {\rm sin}^{1/2} \vartheta 
      {\rm e}^{i \varphi/2} \lbl{14} \\
     && Y_{{1\oo 2},-{1\oo 2}} \propto {\rm sin}^{1/2} \vartheta
      {\rm e}^{- i
    \varphi/2}   \lbl{15}
\ear
With respect to the above set of functions (\ref{14},\ref{15}) 
the angular momentum
operator is indeed not Hermitian. Hence, the set of functions as used by
Merzbacher and Van Winter is indeed not suitable for the representation
of angular operator. But the set (\ref{4}),(\ref{5}) used in the
present paper (and also by
Pandres) is free of such a difficulty and/or inconsistency as discussed by
Merzbacher and Van Winter.

Let me illustrate this on an example. From (\ref{4}) we have for
$l={1\oo 2}$ the following subset ${\cal S}_{1\oo 2}$ of normalized functions:
\bear
     &&Y_{{1\oo 2},{1\oo 2}} = {i\oo \pi} {\rm sin}^{1/2} \vartheta \,
    {\rm e}^{i {\varphi\oo 2}} \lbl{16} \\
    &&Y_{{1\oo 2},-{1\oo 2}} = - {i\oo \pi} {\rm cos} \vartheta \,
 {\rm sin}^{-{1\oo 2}} \vartheta \, {\rm e}^{- {{i \varphi}\oo 2}} \lbl{17}
\ear
If, using (\ref{6}),(\ref{7}), we write
\bear
   &&L_x = {1\oo 2} (L_+ + L_-) \lbl{18} \\
  &&L_y = {1\oo {2i}} (L_+ - L_-) \lbl{19}
\ear
we find after taking into account
\bear
    &&L_+ Y_{{1\oo 2},{1\oo 2}} = 0 \lbl{20a} \\
    &&L_+ Y_{{1\oo 2},-{1\oo 2}} = Y_{{1\oo 2},{1\oo 2}}   \lbl{20} \\
    &&L_- Y_{{1\oo 2},{1\oo 2}} = Y_{{1\oo 2},-{1\oo 2}}  \lbl{21} \\
    &&L_- Y_{{1\oo 2},-{1\oo 2}} = Y_{{1\oo 2},-{3\oo 2}} \lbl{22} \\
    && L_+ Y_{{1\oo 2},-{3\oo 2}} = 0 \lbl{23}
\ear
that the matrix elements of angular momentum operator satisfy:
\bear
   && \langle \mbox{${1\oo 2}{1\oo 2}$}|L_x |\mbox{${1\oo 2}-{1\oo 2}$} 
   \rangle= {1\oo 2} =
    \langle \mbox{${1\oo 2},-{1\oo 2}$}|L_x |\mbox{${1\oo 2}{1\oo 2}$} 
    \rangle^*      \lbl{24} \\
  && \langle \mbox{${1\oo 2}{1\oo 2}$}|L_y |\mbox{${1\oo 2},-{1\oo 2}$}
   \rangle= - {i\oo 2} =
    \langle \mbox{${1\oo 2},-{1\oo 2}$}|L_x |\mbox{${1\oo 2}{1\oo 2}$}
     \rangle^*
     \lbl{27}
\ear
where 
\be
     \langle lm|L_i|lm' \rangle = \int Y_{lm}^* L_i Y_{lm'}\, \dd \Omega\;
     \; \; , \quad \dd \Omega \equiv {\rm sin} \, \vartheta \, \dd \vartheta
     \, \dd \varphi  {\bf }
\lbl{27A} \ee     
Here we have also taken into account that the states with the same $l$ but
different $m$ values are orthogonal. The matrix values
(\ref{24})--(\ref{27}) are just the standard ones. The fact that
$L_- Y_{{1\oo 2},-{1\oo 2}} \neq 0$ has no influence on the values of
matrix elements of angular momentum operator, calculated with respect to the
basis states of ${\cal S}_l$. This is so because of
eq. (\ref{23}). 

In eqs. (\ref{24})--(\ref{27}) we have just the property
that the matrix elements of a Hermitian operator have to satisfy.

Let us now check by explicit integration
whether the operators $L_i$, $i=1,2,3$, satisfy the
requirement for self-adjointness
\be
   (\phi , L_i \psi) = (L_i \phi, \psi) \quad \mbox{for all }
   \phi , \, \psi \in {\cal S}_l 
\lbl{A2}
\ee
with the inner product being defined according to eq.\,(\ref{B5}).
Since any physically admissible $\phi,~\psi$ is by definition a
superposition of
$Y_{lm} \in {\cal S}_l$, it is sufficient to show the relation
(\ref{A2}) for functions $Y_{lm} \in {\cal S}_l$ only. Taking into account
the relations (\ref{18}),(\ref{19}) we find that the condition for
self-adjointnes of the operators $L_i$ becomes
\be
     (Y_{lm},L_+ Y_{lm'}) =  (L_- Y_{lm},Y_{lm'}) \qquad {\rm if~~} m = m'+1
\lbl{P1}
\ee
\be
     (Y_{lm},L_- Y_{lm'}) = (L_+ Y_{lm},Y_{lm'}) \qquad {\rm if~~} m = m'-1
\lbl{P2}
\ee
If we calculate the matrix elements by adopting the usual definition
of the inner product, we have ($m=m'+1$):
\bear
  && \int_0^{2 \pi} \dd \varphi \, \int_0^\pi \dd \vartheta
\,
   {\rm sin} \, \vartheta \,
   Y_{lm}^* \, {\rm e}^{-i \varphi} \left ( -{\p \oo
      {\p \vartheta}} + i \, {\rm cot} \, \vartheta \, 
      {\p \oo {\p \varphi}} \right ) Y_{lm'} \nonumber \\
     && =
 \int_0^{2 \pi} \dd \varphi \, \int_0^\pi \dd \vartheta \, 
 \left [ {\rm e}^{-i \varphi}
    {\p \oo {\p \vartheta}} \left ( Y_{lm}^* \, {\rm sin} \,
    \vartheta \right ) - i \, {\rm cos } \, \vtheta \,
     {\p \oo {\p \varphi}} \left ( 
     Y_{lm}^* \, {\rm e}^{-i \varphi} \right ) \,
     \vartheta \right ] Y_{lm'} + B_{m m'} \nonumber \\
    &&= \int_0^{2 \pi} \dd \varphi \, \int_0^\pi \dd \vartheta \,
   {\rm sin} \, \vartheta \, \left [  {\rm e}^{i \varphi} \left ( {\p \oo
      {\p \vartheta}} + i \, {\rm cot} \, \vartheta \, 
      {\p \oo {\p \varphi}} \right ) Y_{lm} \right ]^*
       Y_{lm'} + B_{mm'}  \lbl{P3}
\ear
The boundary term
\be
B_{mm'}= \int_0^{2 \pi} \dd \varphi \, \int_0^\pi \dd \vartheta \,
   \, \left [-\, {\p \oo {\p \vartheta}}
   \left ( Y_{lm}^*\, {\rm sin} \, \vtheta \, 
   Y_{lm'} \right ) {\rm e}^{-i \vphi} + 
   {\p \oo {\p \vphi}} \left ( Y_{lm}^*\, {\rm e}^{-i \vphi} \,
   Y_{lm'} \right ) {\rm cos} \, \vtheta \right ]  
\lbl{P4}
\ee        
 vanishes if $m > - 3/2$. For instance, if $l=\frac{1}{2},~m={1\oo 2}
 ,~m' = - {1\oo 2}$ the boundary term is equal to   
\be
    - \int_0^{2 \pi} \dd \varphi \, \int_0^\pi \dd \vartheta \,
    \left ( {-i \oo \pi} 
   \right )^2 {\p \oo {\p \vtheta}} \left ( {\rm sin}^{1/2} \, \vtheta
   \, {\rm sin} \, \vtheta \, {\rm sin}^{-1/2} \, \vtheta \, {\rm cos}
   \vtheta \right )   
    = 2 \pi\, \int_0^\pi \, \dd ( {\rm sin} \, \vtheta \,
   {\rm cos} \, \vtheta ) = 0 
\lbl{P5}
\ee
In spite of the fact that the $Y_{{1\oo 2},-{1\oo 2}}$ 
due to ${\rm sin}^{-1/2} \, \vtheta$ is singular at the boundaries
$\vtheta=0,~\vtheta=\pi$, this is compensated by 
${\rm sin}^{1/2} \, \vtheta$ occurring in $Y_{{1\oo 2}{1\oo 2}}^*$, and so
the boundary term is zero. But if $m \le -{3\oo 2}$, the boundary term
does not vanish, because there occurs the product
 ${\rm sin}^m\, \vtheta \, {\rm sin}^{(m+1)}\, \vtheta$ which brings
a singularity.
 
We are going to show that the two illnesses, namely the singularity in
$\int \dd \Omega \, Y_{lm}^* Y_{lm}~,~m \le -{3\oo 2}$, and the singularity
of the boundary term $B_{mm'}$ compensate each other, so that by using the
redefined inner product, the self-adjontness condition (\ref{P1}),(\ref{P2})
are fulfilled for arbitrary $Y_{lm} \in {\cal S}_l$

Let us first illustrate this in the case $l={3\oo 2}$.
Using eq.\,(\ref{4}) we obtain the following four top functions, ( i.e., those
of ${\cal S}_l$):
\bear
    && Y_{{3\oo 2}{3\oo 2}} = K_{3\oo 2} \, \sqrt{6} \, \,{\rm e}^{3i \vphi/2}
     \, {\rm sin}^{3/2}\, \vtheta \nonumber \\
     && Y_{{3\oo 3}{1\oo 2}} = K_{3\oo 2} \, \sqrt{2} \, (-3)\,
      \,{\rm e}^{i \vphi/2} \, {\rm sin}^{1\oo 2}\, \vtheta \, {\rm cos}\, 
      \vtheta \nonumber \\
    && Y_{{3\oo 2},-{1\oo 2}} = K_{3\oo 2} \, \sqrt{{1\oo 2}} \, \, 3 \, 
    {\rm e}^{-i\vphi/2} \, {\rm sin}^{-1/ 2} \, \vtheta \, 
    (2 {\rm cos}^2 \, \vtheta - 1) \nonumber \\
   && Y_{{3\oo 2},-{3\oo 2}} = K_{3\oo 2} \sqrt{{1\oo 6}} \, \,(-3) \, 
   {\rm e}^{-3i \vphi/2} \, {\rm sin}^{-3/2} \, \vtheta \,
   (2 {\rm cos}^2 \, \vtheta - 3)
\lbl{P6}
\ear
where
\be
   K_l \equiv \frac{ (-1)^l}{  \sqrt{2 \pi}}\sqrt{\frac{2 l+1}{2}}
   \frac{1}{2^l \Pi(l)} \qquad
   {\rm end} \qquad K_{{3\oo 2}} = - \, \frac{i \sqrt{2}}{3 \pi}
\lbl{P7}
\ee
Through a direct computation of the explicit action of the operators
$L_+$ and $L_-$ on functions $Y_{lm}$ ddefined in eq.\,(\ref{4}), one finds
that the following relations are satisfied
\bear
    && L_+ Y_{lm} = \sqrt{(l-m)(l+m+1)}\, Y_{l,m+1} \; \;, \quad
    m =l,...,-l, -l-1,l-2,... \lbl{P8} \\
    && L_- Y_{lm} = \sqrt{(l+m)(l-m+1)}\, Y_{l,m-1}\; \; , 
    \quad  m =l,...,-l\; , ~~  m\neq -l \;, \nonumber \\
    && \hs{7.5cm}   m=-l-1,-l-2,...  \lbl{P9}
 \ear    
 
 Using the abbreviation $Y_{{3\oo 2}m} \equiv Y_m$ we now calculate the terms
 in the relation
 \be
     (Y_{-{3\oo 2}},L_-Y_{-{1\oo 2}})_\epsilon = (L_+ Y_{-{3\oo 2}},Y_{-{1\oo
     2}})_\epsilon + B_{-{3\oo 2},-{1\oo 2}} (\epsilon)
\lbl{P8a}
\ee
We obtain
\bear
   &&(Y_{-{3\oo 2}},L_-Y_{-{1\oo 2}})_\epsilon = 
   \sqrt{3}\,  (Y_{-{3\oo 2}},Y_{-{3\oo 2}})_\epsilon = \sqrt{3}\, 
   \int_\epsilon^{\pi-\epsilon} 2 \pi \, \dd \vtheta \, \, 
   {\rm sin}\, \vtheta \, Y_{-{3\oo 2}}^* Y_{-{3\oo 2}} \nonumber \\
   && \hs{2.7cm}= \sqrt{3}\, |K_{{3\oo 2}}|^2 \,2 \pi \,{3\oo 2} (\frac{3
   \pi}{2} - 3 \epsilon -2\, {\rm sin}\,2 \epsilon +{1\oo 4} \,
   {\rm sin}\, 4 \epsilon + 2 \,{\rm cot}\, \epsilon )
\lbl{P8b}
\ear
\bear
 &&(L_+ Y_{-{3\oo 2}},Y_{-{1\oo  2}})_\epsilon  =  
 \sqrt{3}\,  (Y_{-{1\oo 2}},Y_{-{1\oo 2}})_\epsilon = \sqrt{3}\, 
   \int_\epsilon^{\pi-\epsilon} 2 \pi \, \dd \vtheta \, 
   {\rm sin}\, \vtheta \, Y_{-{1\oo 2}}^* Y_{-{1\oo 2}} \nonumber \\
   && \hs{2.7cm}= \sqrt{3}\, |K_{{3\oo 2}}|^2 \,2 \pi \,{9\oo 2} 
   (\frac{\pi}{2} - \epsilon - {1\oo 4} \,
   {\rm sin}\, 4 \epsilon )
\lbl{P10}
\ear
\bear
   B_{-{3\oo 2},-{1\oo 2}} (\epsilon) &=& - 2 \pi \, Y_{-{3\oo 2}}^*Y_{-{1\oo
   2}}\, {\rm sin}\, \vtheta \, \biggl \vert_\epsilon^{\pi-\epsilon}
   {\rm e}^{-i \vphi} \nonumber \\
   &=& \sqrt{3}\, |K_{{3\oo 2}}|^2 \,2 \pi \,(-{3\oo 2})(1 - 4\, {\rm cos}\,
   2\epsilon + {\rm cos}\, 4 \epsilon) \, {\rm cot} \, \epsilon
 \lbl{P11}
 \ear
 from which we can verify that the relation (\ref{P8a}) is indeed
 satisfied for any $\epsilon$. A check is straightforward, if we use
 the symbolic package Mathematica. An easy check by hand can be done
 for small $\epsilon$. We have:
 \bear
     (Y_{-{3\oo 2}},L_-Y_{-{1\oo 2}})_\epsilon &=&
     \sqrt{3}\, |K_{{3\oo 2}}|^2 \,2 \pi \,{3\oo 2}\,
      \left (\frac{3 \pi}{2}+ 2
     \, {\rm cot}\, \epsilon \right) + {\cal O}_1 (\epsilon) \lbl{P12a} \\
(L_+ Y_{-{3\oo 2}},Y_{-{1\oo  2}})_\epsilon &=&  
\sqrt{3}\, |K_{{3\oo 2}}|^2 \,2 \pi \,{3\oo 2}\,
      \left (\frac{3 \pi}{2} \right ) + {\cal O}_2 (\epsilon) 
      \lbl{P12b} \\
  B_{-{3\oo 2},-{1\oo 2}} (\epsilon)&=&    
   \sqrt{3}\, |K_{{3\oo 2}}|^2 \,2 \pi \,{3\oo 2}\,
      \left (- 2
     \, {\rm cot}\, \epsilon \right) + {\cal O}_3 (\epsilon) 
     \lbl{P12c}    
\ear
where ${\cal O}_1,~{\cal O}_2$ and ${\cal O}_3 $ are small $\epsilon$-dependent
terms that go to zero with vanishing $\epsilon$.
We see that the boundary term which grows to infinity, exactly matches the
infinite term in the matrix element
$ (Y_{-{3\oo 2}},L_-Y_{-{1\oo 2}})_{\epsilon \to 0} $ .This enables us to
adopt a suitable renormalization procedure. One possibility is in modifying
the {\it inner product} according to (\ref{B5}) Then eqs.\,(\ref{P1}),(\ref{P2})
written in the form
\be
    \frac{ (Y_{-{3\oo 2}},L_-Y_{-{1\oo 2}})_\epsilon}
    {(Y_{-{3\oo 2}}, Y_{-{3\oo 2}})_\epsilon   }
    = \frac{(L_+ Y_{-{3\oo 2}},Y_{-{1\oo  2}})_\epsilon}
    { (Y_{-{1\oo 2}}, Y_{-{1\oo 2}})_\epsilon  }
\lbl{P13}
\ee
are indeed satisfied, as can be straightforwardly verified.

For arbitrary $l,m$ we have
\be
    (Y_{lm},L_- Y_{l,m+1})_\epsilon = (L_+ Y_{lm},Y_{l,m+1})_\epsilon
     + B_{lm,m+1} (\epsilon) 
\lbl{P14}
\ee
\be
      (Y_{lm},L_- Y_{l,m+1})_\epsilon = (Y_{lm},L_- Y_{l,m+1})_R +
        A_{lm}(\epsilon) = \sqrt{(l-m)(l+m+1)}\, N_{lm} (\epsilon)
        \lbl{P15} \ee
\be       (L_+ Y_{lm},Y_{l,m+1})_\epsilon = (L_+ Y_{lm},Y_{l,m+1})_R +
      A_{l,m+1} (\epsilon) = \sqrt{(l+m)(l-m+1)}\, N_{l,m+1} (\epsilon)
      \lbl{P16}
\ee
Here $R$ denotes the $\epsilon$-independent part which satisfies
\be
   (Y_{lm},L_- Y_{l,m+1})_R = (L_+ Y_{lm},Y_{l,m+1})_R = \sqrt{(l-m)(l+m+1)}
\lbl{P17}
\ee
In the case $l={3\oo 2}$ we read from eqs.\,(\ref{P8b}),(\ref{P16})
that
\be
      (Y_{-{3\oo 2}},L_-Y_{-{1\oo 2}})_R = 
       (L_+ Y_{-{3\oo 2}},Y_{-{1\oo  2}})_R =
        \sqrt{3}\, |K_{{3\oo 2}}|^2 \,2 \pi \, \frac{9 \pi}{4} = \sqrt{3}
\lbl{P18}
\ee
From eqs.\,(\ref{P14})--(\ref{P17}) it follows that the 
$\epsilon$-dependent terms satisfy the relation
\be
    A_{lm}(\epsilon) = A_{l,m+1} (\epsilon) + B_{lm,m+1}(\epsilon)
\lbl{P19}
\ee
Let us now consider the condition for self-adjointness in which the scalar
products are defined according to (\ref{B5}). For arbitrary finite $\epsilon$
we have
\be
     \frac{ (Y_{lm},L_- Y_{l,m+1})_\epsilon   }{N_{lm} (\epsilon)} =
     \frac{(L_+ Y_{lm},Y_{l,m+1})_\epsilon  }{N_{l,m+1} (\epsilon)   }
\lbl{P20}
\ee
Performing the partial integration in the l.h.s. of eq.\,(\ref{P20}), i.e.,
by using eq.\,(\ref{P14}), we obtain
\be
    \frac{(L_+ Y_{lm},Y_{l,m+1})_\epsilon + B_{l,m+1} (\epsilon)  }
    {N_{lm} (\epsilon)} = \frac{(L_+ Y_{lm},Y_{l,m+1})_\epsilon  }
    {N_{l,m+1} (\epsilon)   }
\lbl{P21}
\ee
which in view of eqs.\, (\ref{P15}),(\ref{P16}) becomes
\be
     \sqrt{(l-m)(l+m+1)}\, N_{l,m+1}(\epsilon) + B_{lm,m+1} (\epsilon)
     = \sqrt{(l-m)(l+m+1)}\, N_{lm}(\epsilon)
\lbl{P23}
\ee
Using again (\ref{P15}),(\ref{P16}) we find that the latter relation
is identical to (\ref{P19}). So we have verified that the condition for
self-adjointness (\ref{P20}) holds for arbitrary integration domain,
determined by $\epsilon$, and hence also in the limit $\epsilon \to 0$,
regardless of which $Y_{lm} \in {\cal S}_l$ we take. Thus in the
problematic case of $m \le -{3\oo 2}$, the infinities are regularized
with our definition of the inner product.

This procedure works straightforwardly for physical wave functions
with $m$-values in the range between $l$ and $-l$. It also works for
unphysical wave functions with $m < -l-1$. But there is a problem
at $m=-l-1$. Inserting the latter $m$-value in the self-adjointness
condition (\ref{P20}) we obtain that 
the integral $(L_+ Y_{l,-l-1},Y_{l,-l})_\epsilon$ on the right hand side
does not vary with
$\epsilon$, but is exactly zero, because of the relation
$L_+ Y_{l,-l-1} = 0$ (see eq.\,(\ref{a4})). So the validity of the
condition (\ref{P20}) breaks down in this particular case of the
matrix element between physical and unphysical states.

We have thus shown that angular momentum operator is self-adjoint
with respect to
the domain ${\cal S}_l$ of physical half-integer
spin wave functions, but in general it is not self-adjoint with
respect to the space of all wave functions entering the game.
Since the extra wave functions are unphysical, we need to project
them out.
In the following we will provide an alternative definition of
inner product by which such complications with unphysical wave
functions will be eliminated.

\paragraph{Definition II of inner product -}
The fact that the infinity in a matrix element for $m \le -{3\oo 2}$ 
matches the infinity in the boundary term suggests us to define a
modified inner product in which the infinities are eliminates
\ci{Pandres1}:
\be
    (\psi,\psi') = \int_0^\pi \dd \vtheta\,\left [{\rm sin}\,\vtheta 
    \int_0^{2 \pi} \dd \vphi \, \psi^* \psi' - f(\vtheta) \right ]
\lbl{P24}
\ee
Here $f(\vtheta)$ is a singular function chosen so that the
term in the bracket becomes integrable. For instance, in the case
considered in eq.\,(\ref{P15}), the modified inner product is
$$
   (Y_{lm},L_- Y_{l,m+1}) = \lim_{\epsilon \to 0}\left [ 
   \int_\epsilon^{\pi-\epsilon} \dd \vtheta\,
   \int_0^{2 \pi} \dd \vphi \, Y_{lm}^* L_- Y_{l,m+1} - A_{lm} (\epsilon)
   \right ] = (L_+Y_{lm}, Y_{l,m+1}) $$
\be
    A_{lm} (\epsilon) = 2 \pi \int_{\epsilon}^{\pi - \epsilon} \dd \vtheta
    \, f(\vtheta)
\lbl{P25}
\ee
In particular (see eq.\,(\ref{P8b})) we have
\bear
   (Y_{-{3\oo 2}},L_-Y_{-{1\oo 2}}) &=& 
    \lim_{\epsilon \to 0}\left [ \int_\epsilon^{\pi-\epsilon} \dd \vtheta\,
   \int_0^{2 \pi} \dd \vphi \,Y_{-{3\oo 2}}^*L_- Y_{-{1\oo 2}} \right .
   \nonumber \\
 && \left . ~~~~~~~ -\,\sqrt{3}\, |K_{{3\oo 2}}|^2 \,2 \pi \,
  {3\oo 2}  ( - 3 \epsilon -2\, {\rm sin}\,2 \epsilon +{1\oo 4} \,
   {\rm sin}\, 4 \epsilon + 2 \,{\rm cot}\, \epsilon ) \right ] \nonumber\\
  &=&\sqrt{3}\, |K_{{3\oo 2}}|^2 \,2 \pi \, \frac{9 \pi}{4} =
  \sqrt{3} \nonumber \\
  &=&
 \int_0^\pi \dd \vtheta \, \int_0^{2 \pi} \dd \vphi \,
    (L_+ Y_{-{3\oo 2}})^* Y_{-{1\oo 2}} = (L_+Y_{-{3\oo 2}},Y_{-{1\oo 2}})
\lbl{P26}
\ear
The term $A_{lm} (\epsilon)$ that we added to the integral on the
left hand side of eq.\,(\ref{P26})
is just equal, in the limit $\epsilon \to 0$, to the boundary term
that we obtain after performing the integration per partes
of the right hand side integral.

In fact, the above procedure is a sort of renormalization. The necessity
for such a procedure can be seen from considering, e.g, the matrix
elements $(L_+Y_{-{3\oo 2}},Y_{-{1\oo 2}})$ of eq.\,(\ref{P26}) or
$(Y_{-{1\oo 2}}, L_+ Y_{-{3\oo 2}})$
which are {\it finite} according to
the usual definition of inner product (without a renormalization).
After performing the integration per partes one obtains two infinite
terms which cancel each other, so that the result is still finite,
as it should be.
So it makes sense to redefine the matrix elements $(Y_{-{3\oo 2}},
L_- Y_{-{1\oo 2}})$ and $(L_- Y_{-{1\oo 2}}, Y_{-{3\oo 2}})$ by including
into their definition the corresponding boundary terms.

As observed by Pandres \ci{Pandres1},
the functional $(\psi,\psi')$ satisfies the identities necessary for an inner
product:
$$  (\psi,\psi') \ge 0$$
$$(c \psi, \psi') = c^* (\psi,\psi')$$
$$ (\psi,\psi')^* = (\psi',\psi)$$
\be (\psi + \psi',\psi'') = (\psi,\psi'') + (\psi',\psi'')
\lbl{P27}
\ee
for all $\psi,~\psi'$ spanned by $Y_{lm}$ of eqs.\,(\ref{4}),(\ref{4a})
with $c$ an arbitrary complex constant.

The modified inner product has the following important properties:
\begin{description}
\item{~~(i)} The conditions for self-adjointness of angular momentum
operator  are satisfied for all half-integers $m \le l$.
\item{~(ii)} The basis functions $Y_{lm} \in {\cal S}_l$, i.e., the
{\it physical} ones are orthonormal:
\be
   (Y_{lm},Y_{l'm'}) = \delta_{ll'}\, \delta_{mm'} \; , \qquad m,~m' = l,...,-l
\lbl{P28}
\ee
\item{(iii)} The basis functions $Y_{lm} \in {\cal O}_l,$ i.e., the unphysical
ones, have {\it zero norm}:
\be
    (Y_{lm},Y_{l'm'}) = 0 \; , \qquad m,~m' = -l-1,-l-2,-l-3,...
\lbl{P29}
\ee   
\end{description}
Property (iii) comes from considering the inner product
\be
    (Y_{l,-l-1},Y_{l,-l-1}) \propto (Y_{l,-l-1}, L_- Y_{l,-l}) = 
    (L_+ Y_{l,-l-1},Y_{l,-l}) = 0
\lbl{P30}
\ee
where use has been made of Property (i) and eq.\,(\ref{a3}). In general
\bear
   &&(Y_{l,-l-k},Y_{l,-l-k}) \propto (Y_{l,-l-k},L_- Y_{l,-l-k+1}) \propto
   (L_+ Y_{l,-l-k},Y_{l,-l-k+1}) \nonumber \\
  &&~~~~~~~~~~~~~~~~~~~~ \propto (Y_{l,-l-k+1},Y_{l,-l-k+1}) 
  \propto ...  \propto
    (Y_{l,-l-1},Y_{l,-l-1}) = 0
\lbl{P31}
\ear
which proves Property (iii).
          
{\it Other problems -} Van Winter pointed to a number of problems
and inconsistencies that all can be shown as resulting from his choice
of functions.
 Such problems do not arise with our and Pandres's choice
of functions (\ref{4}),(\ref{5}). Namely, for any function $\psi \in 
{\cal S}_l$ the relations such as
\be
       [L_x , L_y] \psi = i L_z \psi
\lbl{27a}
\ee
\be
      [L_i , \boldsymbol{L}^2] \psi = 0 \; , \qquad L_i = L_x, L_y, L_z
\lbl{27b}
\ee      
\be
     \boldsymbol{L}^2 L_i \psi = l (l+1) L_i \psi
\lbl{27c}
\ee
are valid\footnote{
Crutial here is the relation $L_+ Y_{l,-l-1} = 0$.}.
This is not so for Van Winter's choice of functions.     

However, one problem---discussed by Pauli and Van Winter--- remains even
with our choice of functions, if we take Definition I of inner product.
Namely, a rotation applied to a function
$Y_{lm}$ belonging to ${\cal S}_l$ will give a function outside
${\cal S}_l$. At first sight this seems as an evidence that functions of
${\cal S}_l$ cannot form a representation of rotations and angular momentum.
Following Pandres we will show that this is
not the case, provided that we suitably generalize the concept of
representation space.
With Definition II of inner product no such complication arises,
because the norms of the unphysical states vanish, and consequently
the subspace ${\cal S}_l$ is invariant with respect to rotations.

\subsection{Behaviour of the spherical harmonics with half-integer $l$ values
under rotations}

{\it Behaviour in the presence of Definition I of inner product}

We will now first explore how the spherical functions for half-integer
$l$ values change under infinitesimal rotations. Let a
state\footnote{In order
to simplify the notation we now use the ket notation $|lm \rangle$,
$|\Psi \rangle$, $L_x |\Psi \rangle$, etc., with
the understanding that $\langle \Omega|lm \rangle \equiv Y_{lm}$,
$\langle \Omega|\Psi \rangle \equiv \Psi(\Omega)$, 
$\langle \Omega|L_x|\Psi \rangle =
L_x \Psi(\Omega)$, etc.\,. In fact, we should have used different symbols
for the abstract operator and its representation in the basis $|\Omega \rangle$.
But for simplicity reasons we avoid such complication.}
 $|\Psi \rangle$ with a half-integer of $l$ be a superposition of the states
$|lm \rangle \equiv |m \rangle$ with different values of $m$:
\be
     |\Psi \rangle = \sum_{lm'} |lm' \rangle \langle lm'|\Psi \rangle 
     \; , \qquad m' = l, l-1,..., -l, -l-1, -l-2,...
\lbl{28}
\ee
In general the expansion coefficients $\langle lm'|\Psi \rangle$ are arbitrary.
Let us consider a particular case in which the coefficients are zero for
the values of $m'$ outside ${\cal S}_l$:
\be
     \langle lm'|\Psi \rangle = \left\{ \begin{array}{ll}
                                 {\rm non zero} ~~{\rm if} &m' =
                                  -l,...,l\\   
                                  0  &m' = -l-1,-l-2,...
                                  \end{array}
                                \right.
\lbl{28a}
\ee

Under a rotation around an axis, say $x$-axis, the state changes as
\be
      |\Psi \rangle \rightarrow {\rm e}^{i \epsilon L_x} |\Psi \rangle
\lbl{29}
\ee
where $\epsilon$ is an angle of rotation. For an infinitesimal rotation we have
\bear
       && |\Psi \rangle \rightarrow (1 + i \epsilon L_x) |\Psi \rangle
       \nonumber \\
       && \delta |\Psi \rangle = i \epsilon L_x |\Psi \rangle 
       \lbl{30}
\ear
The projection $\langle m|\Psi \rangle$ changes according to
\be
    \langle l m|\delta|\Psi \rangle \equiv \delta \langle l m|\Psi \rangle =
    i \epsilon \sum_{m'} \langle l m|L_x|m' \rangle \langle l m'|\Psi \rangle
\lbl{31}
\ee

Let us consider the example in which $l={1\oo 2}$, $m={1\oo 2}, -{1\oo 2}$.
Then (\ref{28}) and (\ref{31}) read
\be
    |\Psi \rangle = |\mbox{${1\oo 2}{1\oo 2}$} \rangle  \pbra |\Psi \rangle + 
    \mket \mbra \Psi | \equiv \pket \alpha + \mket \beta 
\lbl{32} 
\ee
\be
    \delta \langle lm|\Psi \rangle = i \epsilon \langle lm|L_x \pket \pbra
    \Psi \rangle + i \epsilon \langle lm|L_x \mket \mbra \Psi \rangle \; ,
    \quad  m = \mbox{$1\oo 2$}, \, -\mbox{$1\oo 2$}
\lbl{33}
\ee
or explicitly
\be
      \delta \pbra \Psi \rangle \equiv \delta \alpha = i \epsilon \pbra 
      L_x \mket \beta = i {\epsilon \oo 2} \beta
\lbl{34}
\ee
\be
      \delta \mbra \Psi \rangle \equiv \delta \beta = i \epsilon \mbra 
      L_x \pket \alpha = i {\epsilon \oo 2} \alpha
\lbl{35}
\ee
where we have taken into account (\ref{18})--(\ref{27}).

Working directly with the functions we have:
\be
   \psi(\Omega) \equiv \langle \Omega |\Psi \rangle = 
   \alpha \psi_{1\oo 2} + \beta
   \psi_{-{1\oo 2}}
\lbl{41a}
\ee
\be
\delta \psi = i \epsilon L_x \psi = i \epsilon L_x \psi_{1\oo 2} \alpha +
i \epsilon L_x \psi_{-{1\oo 2}} \beta
\lbl{41b}
\ee
where $\psi_{1\oo 2} \equiv Y_{{1\oo 2} {1\oo 2}}$ and 
$\psi_{-{1\oo 2}} \equiv Y_{{1\oo 2},-{1\oo 2}}$. Multiplying (\ref{41b})
with $\psi_{-{1\oo 2}}^*$ and $\psi_{{1\oo 2}}^*$, resepectively, and
integrating over $\dd \Omega = {\rm sin} \, \vartheta \,
\dd \vartheta \, \dd \varphi$ we find after taking into
account (\ref{18})-(\ref{27}) that
\be
   \int \dd \Omega \, \psi_{-{1\oo 2}}^* \delta \psi= \delta \beta = i \epsilon 
   \int \dd \Omega \, \psi_{-{1\oo 2}}^* L_x \psi = 
   \int \dd \Omega \, \psi_{-{1\oo 2}}^* L_x \psi_{{1\oo 2}} \alpha =
   {i \oo 2} \epsilon \alpha
\lbl{41c}
\ee
\be
   \int \dd \Omega \, \psi_{{1\oo 2}}^* \delta \psi = 
   \delta \alpha = i \epsilon 
   \int \dd \Omega \, \psi_{{1\oo 2}}^* L_x \psi = 
   \int \dd \Omega \, \psi_{{1\oo 2}}^* L_x \psi_{-{1\oo 2}} \beta =
   {i \oo 2} \epsilon \beta
\lbl{42c}
\ee 
which is the same result as in eqs.( \ref{34}),(\ref{35}). 

The above result demonstrates that under an infinitesimal rotation the expansion
coefficients $\langle lm'|\Psi \rangle$ for $l= {1\oo 2}$,
$m' = \pm {1\oo 2}$ change precisely in the same way as in the usual
theory of spin ${1\oo 2}$
state. From eq.(\ref{31}) we find that this is so in the case of an
arbitrary $l$ and $m' = -l, -l+1,...,l-1,l$ as well.

In a state (\ref{28}) the coefficients $\langle m|\Psi \rangle$ are zero for
the values of $m$ outside ${\cal S}_l$, i.e., for $m > l$ and $m < -l$.
We will now explore how those coefficients change under an infinitesimal
rotation. For the sake of definitness let us again consider the special
case of $l = {1\oo 2}$ and the state given in eq.(\ref{32}) in which
the coefficients $\langle \mbox{$1\oo 2$},- {3\oo 2}|\Psi \rangle , 
\langle \mbox{$1\oo 2$},- {5\oo 2}|\Psi \rangle $, etc., are zero. The change of the
coefficient $\langle - {3\oo 2}|\Psi \rangle$ under the transformation
(\ref{30}) as given by (\ref{31}) and (\ref{33}) reads
\be
     \delta \langle\mbox{$1\oo 2$}, - \mbox{$3\oo 2$} |\Psi \rangle = 
     i \epsilon 
     \langle \mbox{$1\oo 2$}, - \mbox{$3\oo 2$}|
     L_x \mket \mbra \Psi \rangle = {{i \epsilon}\oo 2} \langle
      \mbox{$1\oo 2$}, - \mbox{$3\oo 2$}|
     \mbox{$1\oo 2$}, - \mbox{$3\oo 2$} \rangle \beta 
\lbl{36}
\ee
By rotation we thus obtain a state which is no longer of the form (\ref{32}),
but of the form
\be
      |\Psi' \rangle =  \pket \pbra \Psi' \rangle + \mket \mbra \psi' \rangle
      + |\mbox{$1\oo 2$}, - \mbox{$3\oo 2$} \rangle 
      \langle \mbox{$1\oo 2$}, - \mbox{$3\oo 2$} |\Psi' \rangle
\lbl{37}
\ee
where
\bear
      \pbra \Psi' \rangle &=& \alpha + i {\epsilon \oo 2} \beta 
      \nonumber \\
      \mbra \Psi' \rangle &=& \beta + i {\epsilon \oo 2} \alpha \nonumber \\
      \langle \mbox{$1\oo 2$}, - \mbox{$3\oo 2$} |\Psi' \rangle &=& i {\epsilon \oo 2} 
      \langle -\mbox{$3\oo 2$}|
      \mbox{$1\oo 2$}, - \mbox{$3\oo 2$} \rangle \beta
      \lbl{37a}
\ear
In other words, by a rotation we obtain a state which is outside
${\cal S}_{1\oo 2}$.

What happens if we perform another infinitesimal rotation (\ref{3}) on the
state $|\psi' \rangle$ given in eq.\,(\ref{37}). The coefficients change
according to eq.(\ref{31}) which now read
\bear
     \delta \pbra \Psi' \rangle &=& i \epsilon \pbra L_x \mket
     \mbra \Psi' \rangle +
     i \epsilon \pbra L_x |\mbox{$1\oo 2$}, - \mbox{$3\oo 2$} \rangle
     \langle \mbox{$1\oo 2$}, - \mbox{$3\oo 2$}|\Psi' \rangle  \lbl{38} \\
     \delta \mbra \Psi' \rangle &=& i \epsilon \mbra L_x \pket
     \pbra \Psi' \rangle +
     i \epsilon \mbra L_x |\mbox{$1\oo 2$}, - \mbox{$3\oo 2$} \rangle
     \langle \mbox{$1\oo 2$}, - \mbox{$3\oo 2$}|\Psi' \rangle  \lbl{39} \\
     \delta \langle  \mbox{$1\oo 2$}, - \mbox{$3\oo 2$}|\Psi' \rangle &=& i \epsilon 
     \langle \mbox{$1\oo 2$}, - \mbox{$3\oo 2$}|
     L_x\mket \mbra \Psi' \rangle = i {\epsilon \oo 2} 
     \langle \mbox{$1\oo 2$}, - \mbox{$3\oo 2$}|
     - \mbox{$3\oo 2$} \rangle \mbra \Psi' \rangle \lbl{39a} \\
     \delta \langle  \mbox{$1\oo 2$},-\mbox{$5\oo 2$} |\Psi' \rangle &=& i \epsilon 
     \langle\mbox{$1\oo 2$}, -\mbox{$5\oo 2$}|
     L_x \mbox{$1\oo 2$}, - \mbox{$3\oo 2$} \langle \mbox{$1\oo 2$}, - 
     \mbox{$3\oo 2$}|\Psi' \rangle =
     i {\epsilon\oo 2} \langle 
     \mbox{$1\oo 2$},- \mbox{$5\oo 2$}|\mbox{$1\oo 2$},-\mbox{$5\oo 2$} 
     \rangle 
     \langle \mbox{$1\oo 2$}, - \mbox{$3\oo 2$} |\Psi' \rangle \lbl{39b})
\ear
Writing $L_x$ in terms of $L_+$ and $L_-$ (eq.(\ref{18})) and taking
into  account the relation (\ref{13}) which implies
\be
       L_+ |\mbox{$1\oo 2$},-\mbox{$3\oo 2$} \rangle = 0
\lbl{40}
\ee
we have
\be
       \pbra L_x |\mbox{$1\oo 2$},- \mbox{$3\oo 2$} \rangle = 0 \; \; , \qquad
       \mbra L_x |\mbox{$1\oo 2$},- \mbox{$3\oo 2$} \rangle = 0
\lbl{41}
\ee       
Therefore $\delta \pbra \Psi' \rangle$ and $\delta \mbra \Psi' \rangle$ are
again of the same form as in (\ref{34}),\,(\ref{35}). Becasue of the relation
(\ref{40}), the presence of a non vanishing coefficients 
$\langle\mbox{$1\oo 2$}, - {3\oo 2}|\Psi' \rangle $
has no influence on the transformations of 
$\pbra \Psi' \rangle$ and $\mbra \Psi' \rangle \Psi' \rangle$.

Altogether, our new state after the second rotation is
\be
      |\Psi'' \rangle = \pket \pbra \Psi'' \rangle + 
      \mket \mbra |\Psi'' \rangle
      + |\mbox{$1\oo 2$},-\mbox{$3\oo 2$} \rangle 
      \langle\mbox{$1\oo 2$}, -\mbox{$3\oo 2$} |\Psi'' \rangle +
      |-\mbox{$5\oo 2$} \rangle \langle -\mbox{$5\oo 2$}|\Psi'' \rangle
\lbl{42}
\ee
where
\bear
      \pbra \Psi'' \rangle &=& \pbra \Psi' \rangle + i {\epsilon\oo 2}
      \mbra \Psi' \rangle \lbl{43} \\
       \mbra \Psi'' \rangle &=& \mbra \Psi' \rangle + i {\epsilon\oo 2}
      \pbra \Psi' \rangle \lbl{444} \\
      \langle \mbox{$1\oo 2$},-\mbox{$3\oo 2$}|\Psi'' \rangle &=& i {\epsilon\oo 2} 
      \langle \mbox{$1\oo 2$},-\mbox{$3\oo 2$}|\mbox{$1\oo 2$},
      -\mbox{$3\oo 2$} \rangle \mbra \Psi' \rangle \lbl{45} \\
       \langle \mbox{$1\oo 2$},-\mbox{$5\oo 2$}|\Psi'' \rangle &=& i {\epsilon\oo 2} 
       \langle \mbox{$1\oo 2$},-\mbox{$5\oo 2$}|\mbox{$1\oo 2$},
      -\mbox{$5\oo 2$} \rangle \langle \mbox{$1\oo 2$},-\mbox{$3\oo 2$}| \Psi' 
      \rangle \lbl{46} \\
\ear

Applying now an infinitesimal rotation on $|\Psi'' \rangle$ we find that
\bear
      \delta \pbra \Psi'' \rangle &=& i \epsilon \pbra L_x \mket \mbra \Psi''
      \rangle = i {\epsilon \oo 2} \mbra \Psi'' \rangle \lbl{47} \\
      \delta \mbra \Psi'' \rangle &=& i \epsilon \mbra L_x \mket \pbra \Psi''
      \rangle = i {\epsilon \oo 2} \pbra \Psi'' \rangle \lbl{48}
\ear
which is again a relation of the same form (\ref{34}),\,(\ref{35}) as in the
first and the second infinitesimal rotation.

A rotation brings a state $|\Psi \rangle$ into a state $|\Psi' \rangle$
which lies outside the space spanned, e.g., in the case $l={1\oo 2}$,
by the basis vectors $\pket$, $\mket$,
but the projection $|{\bar \Psi}' \rangle$ onto that space behaves as
the usual spinor. The coefficients $\alpha \equiv \pbra \Psi \rangle$,
$\beta \equiv \mbra \Psi \rangle$ transform into $\alpha' = \pbra
{\bar \Psi}' \rangle = \pbra \Psi' \rangle$, $\beta' \equiv \mbra 
{\bar \Psi}' \rangle = \mbra \psi' \rangle$ in the same way as those of the
usual spinors and their norm is preserved: $|\alpha|^2 +|\beta|^2 = 
|\alpha'|^2 + |\beta'|^2 = 1$.

It is important that under rotation
\be
       |\alpha|^2 +|\beta|^2 = |\alpha'|^2 + |\beta'|^2 = 1
\lbl{49}
\ee
This is essential. In the full space spanned by $\pket, \, \mket , \,
|{1\oo 2},-{3\oo 2} \rangle, \, |{1\oo2},-{5\oo 2} \rangle, ...$,
a vector
$|\Psi \rangle$ transforms in a peculiar way. But its projection
$|{\bar \Psi} \rangle$ onto the space spanned by $\pket,\, \mket$ behaves
as a usual spinor. The matrix elements of $L_x$, $L_y$, $L_z$ in the
states $|lm \rangle = \pket, \, \mket$ are the same as those in the
usual theory of spinors.

Analogous results hold in the case of an arbitrary $l$. 
If we perform an arbitrary succession
of infinitesimal rotations we find that {\it the
coefficients $\langle l m|\Psi \rangle$, $m = -l,...,l$, change under
rotations in the same manner as in the case of spinors.}; this is so because
of the relation (\ref{a3}) which has for a consequence that for $m=-l,..., l$
the matrix elements $\langle lm|L_x|l,-l-1 \rangle$ vanish.
The presence of the
non vanishing coefficients $\langle lm|\psi \rangle$,
$m < -l$, has no influence. The latter coefficients
behave in this respect like ``ghosts". The same is true for a finite rotation
as well, since a finite rotation can be considered as an infinite sequence
of infinitesimal rotations.

Analogous transformations properties hold if we represent states
$|lm \rangle$ by the functions $Z_{lm} = \langle \Omega |lm \rangle$
defined in eq.(\ref{5}). Since there is no reason why just one set of 
the functions, say $Y_{lm}$, should represent spinors, we shall later
consider both sets of functions at once. At the moment let us still keep
on considering the functions $Y_{lm}$ only.

For a {\it finite rotation} $D_R$ a state $|\Psi \rangle$ of ${\cal S}_l$
\be
    |\Psi \rangle = \sum_{m=-l}^l C_m |lm \rangle \quad , \qquad 
    C_m \equiv \langle lm|\Psi \rangle
\lbl{50}
\ee
transforms into another state
\be
     |\Psi' \rangle = D_R |\Psi \rangle = D_R \sum_{m=-l}^l C_m |lm \rangle    
\lbl{51}
\ee
which does no longer belong to ${\cal S}_l$. We can decompose (\ref{51})
according to \ci{Pandres}
\be
    |\Psi' \rangle = D_R |\Psi \rangle  = |{\bar \Psi}' \rangle + |{\cal O} \rangle
\lbl{52}
\ee
where
\be   
       |{\bar \Psi}' \rangle = \sum_{m'=-l}^l C'_{m'} |l m' \rangle
\lbl{53}
\ee
\be
        |{\cal O} \rangle  =\sum_{m'=-l-1,-l-2,...} C'_{m'} |l m' \rangle
\lbl{53a}
\ee
and
\be
    C'_{m'} = \sum_{m=-l}^l \langle l m'|D_R|lm \rangle C_m
\lbl{54}
\ee  
It is important to bear in mind that $|{\cal O} \rangle$ is orthogonal to
$|{\bar \Psi}' \rangle$ :
\be
     \langle {\bar \Psi}'|{\cal O} \rangle = 0
\lbl{55}
\ee
and that
\be
     \sum_{m'=-l}^l |C'_{m'}|^2 = \sum_{m=-l}^l |C_m|^2
\lbl{56}
\ee
Eq.(\ref{53}) can be rewritten as
\be
       |{\bar \Psi}' \rangle   = U |\Psi \rangle
\lbl{57}
\ee
where $U$ is just the usual unitary operator for a rotation of a spinor,
represented by the matrix whose elements are $\langle l m'|D_R|lm \rangle$ :
\be
      U \rightarrow \langle l m'|D_R|lm \rangle \quad m, m' = -l,...,l
\lbl{58}
\ee
Unitarity is assured for all states of ${\cal S}_l$, if one uses
either Definition I, or Definition II of inner product.

A state $|\Psi \rangle$ as given in eq.(\ref{50}) thus transforms under a finite
rotation $D_R$ in such a way that {\it the projection onto the subspace 
${\cal S}_l$ spanned
by the basis vectors $|lm \rangle$, $m=-l,...,l$ is transformed in the
same manner as an ordinary state with half-integer $l$}.
 
The above considerations in eqs. (\ref{50})--(\ref{58}) can be rephrased
by saying that a matrix $D_R$ representing a rotation $R$, calculated in
the basis of functions $Y_{lm}$, has the form
\be
       \begin{pmatrix}
          D_R^{({\cal S}_l)} &\vdots & 0 \\
          \hdotsfor{3}\\
          F_R &\vdots & D_R^{({\cal O}_l)} \\
        \end{pmatrix}
\lbl{c1}
\ee
where $D_R^{({\cal S}_l)}$ is just the usual rotation matrix. 
Whilst the submatrix $D_R^{({\cal S}_l)}$ is hermitian, the
total matrix $D_R$ is not hermitian.
The product
\be
      D_{RS} = D_R D_S= \begin{pmatrix}
          D_R^{({\cal S}_l)} D_S^{({\cal S}_l)} &\vdots& 0 \\
          \hdotsfor{3}\\
          F_R  D_S^{({\cal S}_l)}+ D_R^{{(\cal O)}_l} F_S &\vdots & D_R^{({\cal O}_l)}  
          D_S^{({\cal O}_l)}
          \end{pmatrix}
 \lbl{c2}
 \ee 
has the same form as (\ref{c1}). The matrices $D_{RS}^{({\cal S}_l)}=
D_R^{({\cal
S}_l)} D_S^{({\cal S}_l)} $ and $D_{RS}^{({\cal O}_l)}=D_R^{({\cal
S}_l)} D_S^{({\cal O}_l)} $ provide us, respectively, with an $l(l+1)$-dimensional
and infinite dimensional representation of the rotation group.The original
representation $D_R$
is thus reducible. The representation space is split into two subspaces:
the $l(l+1)$-dimensional space ${\cal S}_l$ and the infinite dimensional
space ${\cal O}_l$. From the form of the matrix (\ref{c1}) we find that the
subspace ${\cal O}_l$ is invariant, whilst the subspace ${{\cal S}_l}$ is not
invariant.
 
 A representation is said to be fully reducible\footnote{See, e.g., ref.
\ci{Hamermesh} }, if both subspaces
are invariant, i.e., if it is possible to find a basis in which $F_R=0$.
In many important cases this happens to be the case. But the representation
with the basis given in terms of spherical harmonics is not fully reducible
(in the above sense) for half-integer $l$-values. And yet, according to
the ordinary representation theory, it is reducible, since $D_R^{({\cal S}_l)}$
and $D_R^{({\cal O}_l)}$ in eq.\,(\ref{c1}) are in themselves representations
of the 3-dimensional rotation group. We thus see that the spherical harmonics
with $l={1\oo 2},{3\oo 2},{5\oo 2},...,$ do fit into the theory of group
representations, only the subspace ${{\cal S}_l}$ is not invariant. If initially
we have a state
\be 
      \begin{pmatrix} C \\ ... \cr 0 \end{pmatrix} \in {{\cal S}_l}
\; ,
      \quad C \equiv C_m = \langle lm| \psi \rangle
\lbl{c3}
\ee
then after applying a rotation, e.g., once and twice, we have respectively
\be
     D_S   \begin{pmatrix}C \\ ... \\ 0 \end{pmatrix}= 
            \begin{pmatrix} D_S^{({\cal S}_l)} C\\
                    \hdotsfor{1}\\
            F_S\, C \end{pmatrix}  \; , \quad    D_R D_S   
           \begin{pmatrix}C \\ ...\\ 0  \end{pmatrix}= 
           \begin{pmatrix}D_R^{({\cal S}_l)} D_S^{({\cal S}_l)} C\cr
            ............................... \cr
          (F_S D_S^{({\cal S}_l)} + D_R^{({\cal O}_l)} F_S)\, C \end{pmatrix}
\lbl{c4}
\ee 
Since ${D_R^{({\cal S}_l)}},~ D_S^{({\cal S}_l})$ are just the ordinary
rotation matrices,
the states
$C,~ D_S^{({\cal S}_l)} C,~ D_R^{({\cal S}_l)} D_S^{({\cal S}_l)} C$ in the
subspace ${{\cal S}_l}$ are normalized according to (\ref{56}) and they behave as
ordinary half-integer $l$ states. Analogous considerations hold for
the basis functions $Z_{lm}$.

{\it Behaviour in the presence of the Definition II of scalar
product}

If we adopt the Definition II of inner product, then the situation simplifies
significantly because of the validity of eqs.\, (\ref{P28}),(\ref{P29})
which say that the physical functions are orthonormal, whereas the
unphysical functions have vanishing norms. Consequently, the
matrix elements of the angular momentum operator between such
unphysical states vanish. The complications described in eqs.\,
(\ref{36})--(\ref{39b}) do not arise. A rotation does not bring
a state, initially in ${\cal S}_l$, into a state with components
outside ${\cal S}_l$. The latter space is now invariant.
Therefore, a residual vector
$|{\cal O} \rangle$ that occurred under a rotationin eq.\,(\ref{52}),
does no longer occur. A matrix $D_R$
representing a rotation has nonvanishing elements only between
the physical functions. That is, in eq.\,(\ref{c1}) the submatrix 
$D_R^{({\cal S}_l)}$ is different from zero, whilst $F_R$ and
$D_R^{({\cal O}_l)}$ are zero. Thus the the spherical harmonics with
half-integer spin values form a fully reducible representation
of the 3-dimensional rotation group.

\subsection{Inclusion of the functions $Z_{lm}$ into the description of
half-integer spin}
 
If one looks at the functions $Y_{{1\oo 2}{1\oo 2}}$ and 
$Y_{{1\oo 2},-{1\oo 2}}$ (eqs.(\ref{16}),(\ref{17})) one finds
that they have completely different
forms, so that they cannot be related by a
transformation such as a rotation, space reflection or time
reversal. On the other hand, we would expect that a state which has
only its spin direction reversed should be obtained from the original
state by any of those transformations. Does it means that $Y_{lm}$ for
half integer $l$ are not suitable for the description of spin half states
after all? Such a conclusion would be too hasty, since besides the functions 
$Y_{lm}$ there are also the functions $Z_{lm}$ given in eq.(\ref{5}).
In particular, for $l={1\oo 2}$, we have
\be
     Z_{{1\oo 2}{1\oo 2}} = {1\oo \pi} {\rm sin}^{-1/2} \, \vartheta \,
     {\rm cos} \, \vartheta \, {\rm e}^{i \varphi \oo 2}
\lbl{59}
\ee
\be
    Z_{{1\oo 2},-{1\oo 2}} =  {1\oo \pi} {\rm sin}^{1/2} \, \vartheta \,
    {\rm e}^{-{i \varphi \oo 2}} \hs {1cm}
\lbl{60}
\ee
They satisfy the following relations
\be
      L_- Z_{{1\oo 2}{1\oo 2}} = Z_{{1\oo 2},-{1\oo 2}}
\lbl{60a}
\ee
\be
   L_+ Z_{{1\oo 2},-{1\oo 2}} = Z_{{1\oo 2}{1\oo 2}} \hs{6mm}
\lbl{60b}
\ee

A state $|lm \rangle$ with half integer $l$ can be represented either by
functions $Y_{lm}$ or $Z_{lm}$, or, in general, by a superposition
\be
       |lm \rangle \rightarrow \Psi_{lm} = a Y_{lm} + b Z_{lm}
\lbl{61}
\ee
where $a$, $b$ are complex constants, such that $|a|^2 + |b|^2 = 1$.

For $l={1\oo 2}$ eq. (\ref{61}) becomes
\bear
    |\mbox{${1\oo 2}{1\oo 2}$} \rangle \rightarrow
     \psi_{{1\oo 2}{1\oo 2}} &=&
     \left ( a {i\oo \pi} {\rm sin}^{1/2} \, \vartheta  +
    b {1\oo \pi} {\rm sin}^{-1/2} \, \vartheta \, {\rm cos}\, 
    \vartheta \right )
    {\rm e}^{{i \varphi}\oo 2} \lbl{62} \\
    |\mbox{${1\oo 2},-{1\oo 2}$} \rangle
     \rightarrow \psi_{{1\oo 2},-{1\oo 2}} &=&
    L_- \psi_{{1\oo 2}{1\oo 2}} = 
     \left (- a{i\oo \pi} \, {\rm sin}^{-1/2} \, \vartheta \, 
   {\rm cos} \, \vartheta  - b \,{1\oo \pi} {\rm sin}^{1/2} \, 
   \vartheta \right ) 
     {\rm e}^{{-i \varphi}\oo 2}
\lbl{63}
\ear
The preceding expressions demonstrate that functions
 $\psi_{{1\oo 2}{1\oo 2}}$
and $\psi_{{1\oo 2},-{1\oo 2}}$ are given by similar expressions. There is
no longer such a drastic difference between the $m=1/2$ and $m=-1/2$
functions. Functions $\psi_{{1\oo 2}{1\oo 2}},~\psi_{{1\oo 2},-{1\oo 2}}$
are eigenfunctions of $\boldsymbol{L}^2$ and $L_z$ with the eigenvalues
$l(l+1) = 3/4$ and $m = {1\oo 2},~-{1\oo 2}$, respectively.

Let us now study the behaviour of the functions (\ref{62}),
(\ref{63}) under some
transformations of particular interest.

 a) {\it $180^0$ rotation around the $y$-axis}. 
The polar coordinates change according to
\bear
     && r \rightarrow  r  \nonumber \\
     && \vartheta \rightarrow \pi - \vartheta  \lbl{65} \\
     && \varphi \rightarrow \pi - \varphi \nonumber 
\ear
This gives
\bear
    && x = r\, {\rm sin} \, \vartheta \, {\rm cos} \, 
    \varphi \rightarrow 
    r \, {\rm sin} \, (\pi -\vartheta) \, {\rm cos} \, (\pi -\varphi) 
    = - x \nonumber  \\
    && y = r \, {\rm sin} \, \vartheta \, {\rm sin} \varphi \rightarrow 
     r \, {\rm sin} \, (\pi -\vartheta) \, {\rm sin} (\pi -\varphi)
      = y \lbl{66} \\
    && z = r \, {\rm cos} \, \vartheta \rightarrow  
    r \, {\rm cos} \, (\pi -\vartheta) = - z
    \nonumber
\ear
where we have taken into account ${\rm sin} (\pi - \vtheta) = 
{\rm sin} \, \vtheta,~ {\rm cos} (\pi-\vphi) = - {\rm cos} \, \vphi$.

Under the change of coordinates (\ref{65}),(\ref{66}), the basis functions
transform as
\bear
   &&Y_{{1\oo 2}{1\oo 2}} (\vtheta,\vphi) \rightarrow {i\oo \pi} \, 
   {\rm sin}^{1/2} (\pi - \vtheta) \, {\rm e}^{{i\oo 2} (\pi -\vphi)}
   = - Z_{{1\oo 2},-{1\oo 2}} (\vtheta,\vphi) \nonumber \\
    &&Z_{{1\oo 2}{1\oo 2}} (\vtheta,\vphi) \rightarrow {1\oo \pi} \,
    {\rm sin}^{-1/2} \,(\pi -\vtheta)\, {\rm cos} \, (\pi-\vtheta) \,
    {\rm e}^{{i\oo 2} (\pi-\vtheta)} = Y_{{1\oo 2},-{1\oo 2}}(\vtheta,\vphi)
\lbl{d1}
\ear
so that 
\be
     \psi_{{1\oo 2}{1\oo 2}}  = a Y_{{1\oo 2}{1\oo 2}} + 
     b Z_{{1\oo 2}{1\oo 2}} \rightarrow {\psi'}_{{1\oo 2}{1\oo 2}} = 
     - a Z_{{1\oo 2},-{1\oo 2}} + b Y_{{1\oo 2},-{1\oo 2}} 
\lbl{d2}
\ee
Comparing the transformed wave function ${\psi'}_{{1\oo 2}{1\oo 2}}$
with
\be
    \psi_{{1\oo 2},-{1\oo 2}} = a Y_{{1\oo 2},-{1\oo 2}} 
    + b Z_{{1\oo 2},-{1\oo 2}}
\lbl{d2a}
\ee    
we find that they are related according to
\be
    {\psi'}_{{1\oo 2}{1\oo 2}} = A \psi_{{1\oo 2},-{1\oo 2}} =
    {\tilde \psi}_{{1\oo 2},-{1\oo 2}}
\lbl{d3}
\ee
Here $A$ denotes the transformations which changes $a$ into $b$ and 
$b$ into $-a$.
We see that under the $180^0$ rotation around the $y$-axis the function
$\psi_{{1\oo 2}{1\oo 2}} (\vartheta, \varphi)$ becomes the function
$\psi_{{1\oo 2},-{1\oo 2}} (\vartheta, \varphi)$, apart from an active
SU(2) ``rotation"\footnote{The existence of an SU(2) transformation in the
space spanned by $Y_{lm},~Z_{lm}$ was previously discussed
by Pandres \ci{Pandres}.}
in the space spanned by the basis functions $Y_{{1\oo 2},-{1\oo 2}}$
and $Z_{{1\oo 2},-{1\oo 2}}$. In other words, the $180^0$ rotation of
the coordinate axes (\ref{66}) transforms the function
$\psi_{{1\oo 2}{1\oo 2}}$ into the function which is of the same form as the
function $\psi_{{1\oo 2},-{1\oo 2}}$ (see eq.(\ref{63})), only the
coefficients are different. They are changed by an SU(2) transformation which
in matrix form reads
\be    
         \begin{pmatrix} ~0 & 1 \cr
                   -1 & 0 \end{pmatrix} \begin{pmatrix} a \cr b \end{pmatrix}
       = \begin{pmatrix} ~b \cr -a \end{pmatrix}
\lbl{d4}
\ee
where
\be
       \psi_{{1\oo 2},-{1\oo 2}} \rightarrow \begin{pmatrix} a \cr b 
       \end{pmatrix} \; , 
       \quad  {\tilde \psi}_{{1\oo 2},-{1\oo 2}} \rightarrow
        \begin{pmatrix} ~b \cr -a \end{pmatrix}  
\lbl{d5}
\ee
The function   ${\psi'}_{{1\oo 2}{1\oo 2}}$ which we obtain from
$\psi_{{1\oo 2}{1\oo 2}}$ by the change of coordinates (\ref{65}),(\ref{66})
is an eigenfunction of $\boldsymbol{L}^2$ and $L_z$ with the eigenvalues
$l(l+1) = 3/4$ and $m = - 1/2$, respectively. Therefore we may write
${\tilde \psi}_{{1\oo 2},-{1\oo 2}}$ instead of ${\psi'}_{{1\oo 2}{1\oo 2}}$.

Let us consider two particular cases of special interest:

{\it Case I}
\be
    a = {1\oo \sqrt{2}} \; , \quad b = {i\oo \sqrt{2}}
\lbl{d5a}
\ee
Then eq. (\ref{d4}) gives
\be    
         \begin{pmatrix} ~0 & 1 \cr
                   -1 & 0 \end{pmatrix} {1\oo \sqrt{2}} 
                   \begin{pmatrix} 1 \cr i \end{pmatrix}
         = {i\oo \sqrt{2}} \begin{pmatrix} 1 \cr i \end{pmatrix}
\lbl{d6}
\ee
i.e.,
\be
     {\tilde \psi}_{{1\oo 2},-{1\oo 2}}  = i \psi_{{1\oo 2},-{1\oo 2}} 
\lbl{d7}
\ee
We see that the particular wave function
\be
     \chi_{{1\oo 2}{1\oo 2}} = {1\oo \sqrt{2}} (Y_{{1\oo 2}{1\oo 2}} 
     + i Z_{{1\oo 2}{1\oo 2}} )
\lbl{d8a}
\ee
for which we introduce the new symbol $\chi$
transforms under the $180^0$ rotation around $y$-axis into the wave
function
\be
 \chi'_{{1\oo 2}{1\oo 2}} ={\tilde \chi}_{{1\oo 2},-{1\oo 2}}
 = {i\oo \sqrt{2}} (Y_{{1\oo 2},-{1\oo 2}} + i Z_{{1\oo 2},-{1\oo 2}})
 = i \chi_{{1\oo 2},-{1\oo 2}}
\lbl{d8b}
\ee
which is equal to the wave function $\chi_{{1\oo 2},-{1\oo 2}}$
multiplied by $i$.

{\it Case II}
\be
    a= {i\oo \sqrt{2}} \; , \quad b = {1\oo \sqrt{2}}
\lbl{dd1}
\ee
Then the particular function is
\be
    \theta_{{1\oo 2}{1\oo 2}}= {1\oo \sqrt{2}} (i Y_{{1\oo 2}{1\oo 2}}
    + Z_{{1\oo 2}{1\oo 2}} )
\lbl{dd2}
\ee
and it transforms under the rotation (\ref{65}) into
\be
    \theta'_{{1\oo 2}{1\oo 2}} = {\tilde \theta}_{{1\oo 2},-{1\oo 2}}
    = {i\oo \sqrt{2}} (Y_{{1\oo 2},-{1\oo 2}} + i Z_{{1\oo 2},-{1\oo 2}} )
    = - i \theta_{{1\oo 2},-{1\oo 2}} 
\lbl{dd3}
\ee

To sum up, for the particular choice of coefficients (\ref{d5a}) and (\ref{dd1})
(Case I and Case II) the wave functions transform under the $180^0$
rotation (\ref{65}) according to
\be
    \chi_{{1\oo 2}{1\oo 2}} \rightarrow  i \chi_{{1\oo 2},-{1\oo 2}}
\lbl{dd4}
\ee
\be
    \theta_{{1\oo 2}{1\oo 2}} \rightarrow - i \theta_{{1\oo 2},-{1\oo 2}}
\lbl{dd5}
\ee

b) {\it Reflection of $x$-axis}
\be
   \begin{array}{l}
   x \rightarrow -x \\
   y \rightarrow ~ y \\
   z \rightarrow ~ z
\end{array}   \qquad \mbox{or} \qquad
\begin{array}{l}
     r \rightarrow  r   \\
     \vartheta \rightarrow \vartheta   \\
     \varphi \rightarrow \pi - \varphi  
\end{array}
\lbl{d10}
\ee
This gives
\bear
   &&Y_{{1\oo 2}{1\oo 2}} (\vtheta,\vphi)
    \rightarrow {i \oo \pi} \, {\rm sin}^{1/2} \,
   \vtheta \, {\rm e}^{{i \oo \pi} (\pi - \vphi)} 
   = - Z_{{1\oo 2},-{1\oo 2}} (\vtheta,\vphi)
   \nonumber \\
   &&Z_{{1\oo 2}{1\oo 2}}(\vtheta,\vphi) 
   \rightarrow {1\oo \pi} {\rm sin}^{-1/2} \,
   \vtheta \, {\rm cos} \, \vtheta \, {\rm e}^{{i \oo \pi} (\pi - \vphi)}
   = - Y_{{1\oo 2},-{1\oo 2}}(\vtheta,\vphi) \lbl{d11}
\ear
The transformation of a generic wave function reads
\be
   \psi_{{1\oo 2}{1\oo 2}} = a Y_{{1\oo 2}{1\oo 2}} + b Z_{{1\oo 2}{1\oo 2}}
   \rightarrow {\psi'}_{{1\oo 2}{1\oo 2}} = - a Z_{{1\oo 2},-{1\oo 2}}
   - b Y_{{1\oo 2},-{1\oo 2}} = {\tilde \psi}_{{1\oo 2},-{1\oo 2}}
\lbl{d12}
\ee
Comparing ${\psi'}_{{1\oo 2}{1\oo 2}}
={\tilde \psi}_{{1\oo 2},-{1\oo 2}}$ with 
$\psi_{{1\oo 2},-{1\oo 2}}$ we have
\be
           \begin{pmatrix} ~0 & - 1 \cr
                   -1 & ~0 \end{pmatrix} \begin{pmatrix} a \cr b 
                   \end{pmatrix}  
          =  \begin{pmatrix} -b \cr -a \end{pmatrix}
\lbl{d13}
\ee
where
\be
    \psi_{{1\oo 2},-{1\oo 2}} \rightarrow \begin{pmatrix} a \cr b 
    \end{pmatrix}
    \; , \qquad {\tilde \psi}_{{1\oo 2},-{1\oo 2}}
    \rightarrow \begin{pmatrix} -b \cr -a \end{pmatrix}
\lbl{d13a}
\ee

For the particualr choice of coefficients, (\ref{d4}) and (\ref{dd1})
(Case I and Case II), we find that under the reflection (\ref{d10})
the corresponding wave functions transform according to
\be
    \chi_{{1\oo 2}{1\oo 2}} \rightarrow - \theta_{{1\oo 2},-{1\oo 2}}
\lbl{dd4a}
\ee
\be
    \theta_{{1\oo 2}{1\oo 2}} \rightarrow - \chi_{{1\oo 2},-{1\oo 2}}
\lbl{dd5a}
\ee
We see that the reflection interchanges functions $\chi$ and $\theta$.

c) {\it Space inversion}
\be
   \begin{array}{l}
   x \rightarrow -x \\
   y \rightarrow ~ -y \\
   z \rightarrow ~ -z
\end{array}   \qquad \mbox{or} \qquad
\begin{array}{l}
     r \rightarrow  r   \\
     \vartheta \rightarrow \vartheta -\pi  \\
     \varphi \rightarrow \varphi + \pi
\end{array}
\lbl{dd6}
\ee
We then find
\be
    \chi_{{1\oo 2}{1\oo 2}} \rightarrow  \theta_{{1\oo 2}{1\oo 2}}
\lbl{dd6a}
\ee
\be
    \theta_{{1\oo 2}{1\oo 2}} \rightarrow - \chi_{{1\oo 2}{1\oo 2}}
\lbl{dd7}
\ee

The particular wave functions $\chi_{{1\oo 2}{1\oo 2}}$ and 
$\theta_{{1\oo 2}{1\oo 2}}$ have good behaviour under the $180^0$ rotation
(\ref{65}), because $\chi_{{1\oo 2}{1\oo 2}}$ transforms into
$\chi_{{1\oo 2},-{1\oo 2}}$ and $\theta_{{1\oo 2}{1\oo 2}}$ into
$\theta_{{1\oo 2},-{1\oo 2}}$, apart from a factor $i$ or $-i$. Neglecting
the latter factor, the rotated state is distinguished from the original
state in the sign of the quantum number $m=\pm {1\oo 2}$.
The situation is different in the case of space reflection and space inversion.
The latter transformations interchange the type of the wave function.

Inspecting now {\it rotations of coordinates axes} for other angles, e.g.,
the $2 \pi$ rotation in the $(x,z)$-plane, we find that it brings
$\chi_{{1\oo 2}{1\oo 2}} \rightarrow \chi_{{1\oo 2}{1\oo 2}}$ and
$\theta_{{1\oo 2}{1\oo 2}} \rightarrow \theta_{{1\oo 2}{1\oo 2}}$. That is,
a $2 \pi$ rotation of coordinates axes transforms a wave function
$\chi_{{1\oo 2}{1\oo 2}}$ or $\theta_{{1\oo 2}{1\oo 2}}$ back into
the original  wave function. This is so because a rotation of a
coordinate frame does not affect a physical system in question; from
the point of view of the physical system it is a {\it passive transformation}.
Consider now the popular illustration of spinor by means of
a ball connected to a box with elastic threads\footnote{The box may
represent the entire environment.}. The  {\it active
transformation} of the latter system, corresponding to the rotation of
a coordinate frame, is a rotation of the box\footnote{If the box represents
the environment, then the environment together with the attached ball
is rotated.} that keeps the relative
orientation of the ball unchanged. The box together with the ball is rotated.
On the contrary, rotations by which one illustrates
the spinor properties affect the ball only. A $2 \pi$ rotation of the ball
then entangles the ball and the box in such a way that the transformed
system is not equivalent to the original system. A $4 \pi$ rotation
is needed in order to bring the system back into its original state.

The above considerations demonstrate the general rule that for wave functions
which
 represent spinors, a relation such as $D_R \psi (x) = 
\psi (R^{-1} x )$,
 valid for {\it scalars},
does not hold. {\it Wave functions representing spinors do not transform
as scalars.}
 Here $D_R$ is a linear operator which acts on functions
$\psi (x)$, whilst $R$ is a rotation which acts on coordinates $x$. 
In particular, $R$ can be a rotation around $y$-axis and
$D_R = {\rm exp} (i \alpha L_y)$, i.e.,
the operator analogous to the one considered in eq.(\ref{29}). 
The case of $R$ for $\alpha = \pi$ (i.e., $180^0$) has been considered in
eqs.(\ref{65})--{\ref{d8b}).

Inclusion of wave functions which do not behave as scalars under rotations,
is one amendment to the notion of representation space.
Functions that can form a representation of the
3-dimensional rotation group need not be scalars. This is in agreement with
the fact that spinors are indeed not scalars.

If we take  into account also the states represented by functions $Z_{lm}$,
so that the basis is given in terms of functions 
\be
  \psi_{lm} = a Y_{lm}+b Z_{lm} = A \, \chi_{lm} + B \, \theta_{lm}
\lbl{D00}
\ee
spanning a space ${\cal S}_l$,
and compute the matrix elements
\be
   \langle lm|D_R|lm' \rangle = \int \dd \Omega \, \psi_{lm}^* \, 
   e^{i \boldsymbol{L} \boldsymbol{\alpha} \psi_{lm'}}
\lbl{D0}
\ee
we find that the transformation matrix representing a rotation $R$ has the
form
\be
    D_R = \begin{pmatrix} D_R^{({\cal O}_l^+)} &\vdots& G_R &\vdots & 0 \cr
    \hdotsfor{5}\cr
    0&\vdots & D_R^{({\cal S}_l)}& \vdots& 0 \cr
    \hdotsfor{5} \cr
    0& \vdots& F_R & \vdots & D_R^{({\cal O}_l^-)}
    \end{pmatrix}
    \lbl{D1}
\ee
Now ${\cal S}_l$ denotes a space spanned by the 
functions $\psi_{lm}$ for $m=-l,...,l$ (i.e., superpositions given in
eq.\,(\ref{D00})),   while ${\cal O}_l^-$ and ${\cal O}_l^+$ are the corresponding
spaces for $m<-l$ and $m>l$, respectively.

The product of two rotations gives
\be
D_{RS} = D_R D_S = \begin{pmatrix} D_R^{({\cal O}_l^+)} D_S^{({\cal O}_l^+)}
  &\vdots&
 D_R^{({\cal O}_l^+)} G_R + G_R D_S^{({\cal S}_l)} &\vdots & 0 \cr
    \hdotsfor{5}\cr
    0&\vdots & D_R^{({\cal S}_l)} D_S^{{(\cal S}_l)} & \vdots& 0 \cr
    \hdotsfor{5}\cr
    0& \vdots& F_R D_S^{{(\cal S}_l)}+  D_R^{({\cal S}_l)}F_S & \vdots & 
    D_R^{({\cal O}_l^-)}
D_S^{({\cal O}_l^-)}
    \end{pmatrix}
    \lbl{D2}
    \ee
which is of the same form as (\ref{D1}): $ D_R^{({\cal O}_l^+)}$, 
$ D_R^{({\cal S}_l)}$ and $ D_R^{({\cal O}_l^-)}$ are in themselves representations
of rotations.

Suppose that initially we have a state vector
\be
    \begin{pmatrix}0\cr
            ...\cr
            C \cr
            ...\cr
            0 \end{pmatrix} \in {\cal S}_l \;  ,\quad C= C_{lm} = 
            \langle lm | \psi \rangle
            \; , \; \; m= -l,...,l
\lbl{D3}
\ee
then after applying a rotation, e.g., once and twice, we have
\be
   D_S  \begin{pmatrix}0\cr
            ...\cr
            C \cr
            ...\cr
            0 \end{pmatrix} =
    \begin{pmatrix}G_S C\cr
            ..........\cr
            D_S^{({\cal S}_l)} C \cr
            ..........\cr
            F_S C \end{pmatrix} \; ,\quad  D_R D_S \begin{pmatrix}0\cr
            ...\cr
            C \cr
            ...\cr
            0 \end{pmatrix}=
      \begin{pmatrix}(D_R^{({\cal O}_l^+)} G_S + G_R D_S^{({\cal S}_l)}  ) C\cr
            ....................................\cr
            D_R^{({\cal S}_l)} D_S^{({\cal S}_l)} C \cr
            ....................................\cr
          ( F_R  D_S^{({\cal S}_l)}+ D_R^{({\cal O}_l^-)} F_S) C \end{pmatrix}
\lbl{D4}
\ee

Again $ D_R^{({\cal S}_l)}$ and  $D_S^{({\cal S}_l)}$ are the ordinary, unitary,
rotation
matrices for half integer values of $l$. They act in the subspace ${\cal
S}_l$. Although the latter subspace is not invariant under rotations, it
holds that the norms of the states $C\in {\cal S}_l$, and the corresponding
rotated states
$~D_S^{({\cal S}_l)}C $, $~ D_R^{({\cal S}_l)} D_S^{({\cal S}_l)} C$
are invariant.

This is another amendment to the notion of ``forming a representation'':
a representation
(sub)space ${\cal S}_l$ need not be invariant, provided that the norms of
the states  projected into ${\cal S}_l$ are preserved under the action of the
group elements.

The above considerations are valid for Definition I of inner product.
With Definition II the situation again simplifies significantly,
since in matrix $D_R$ of eq.\,(\ref{D1}) only the piece
$D_R^{{\cal S}_l}$ is different from zero, whilst other pieces
all vanish.          

We have seen that the spherical harmonics with half-integer values of $l$ can
represent the states with half-integer values of angular momentum.
{\it But this cannot be the states of orbital angular momentum, since it is
well known experimentally that orbital angular momentum can have 
integer values only}\footnote{A reason of why to reject $Y_{lm}$ and $Z_{lm}$
with half-integer $l$-values in the description of orbital angular
momentum was given correctly by Dirac \ci{DiracBook}. In the free case, a complete
set of solutions to the Schr\" odinger equation consists of plane waves,
which are single valued. The latter property has to be preserved when we use
another representation, i.e., one with spherical harmonics. (See also
Sec.\,3.2.)}.
{\it Hence, our Schr\" odinger basis for spinor representation of the
3-dimensional rotation group cannot refer to the ordinary configuration
space of positions, but to an internal space associated with
every point of the ordinary space.} A possible internal space is the space
of particle's velocities, or equivalently, of accelerations.
This will be discussed in Sec.\,3.

\paragraph{On the SU(2) in the space spanned by functions $\chi_{lm}$ and
$\theta_{lm}$\,.}

Functions $\chi_{lm}$ and $\theta_{lm}$ are linearly independent.
Let us assume that 
for fixed $l$, $m$ they represent two distinct quantum states classified
by eigenvalues of an operator $T_3$. Let us denote those states as\
\ci{Pandres}
\be
       |lm \Lambda \rangle  \; \; ,
       \qquad \Lambda = \mbox{${1\oo 2}$} , ~-\mbox{${1\oo 2}$}
\lbl{68b}
\ee
so that
\be
    \langle \Omega |lm,\mbox{${1\oo 2}$} \rangle = \chi_{lm} \quad {\rm and} 
    \quad
     \langle \Omega |lm,-\mbox{${1\oo 2}$} \rangle = \theta_{lm}
\lbl{68c}
\ee
The operator $T_3$ is defined by
\be
     T_3 |lm \Lambda \rangle = \Lambda |lm \Lambda \rangle 
\lbl{69}
\ee
We can also define the operators $T_1$ and $T_2$ so that $T^{\pm}
= T_1 \pm i T_2$ connect the states with different values of $\Lambda$
\bear
     &&T^+  |lm,\mbox{$-{1\oo 2}$} \rangle =
     |lm,\mbox{${1\oo 2}$} \rangle  \nonumber \\
     &&T^-  |lm,\mbox{${1\oo 2}$} \rangle =
     |lm,\mbox{$-{1\oo 2}$} \rangle  \nonumber \\
     &&T^+  |lm,\mbox{${1\oo 2}$} \rangle = 0 \quad , \qquad
     T^-  |lm,\mbox{$-{1\oo 2}$} \rangle = 0
\lbl{70}
\ear
The matrices which represent $T_\alpha$, $\alpha = 1,2,3$ on the basis
$|lm \Lambda \rangle $ are just the Pauli matrices. $T_\alpha$ are the
generators of the group SU(2) and they commute with the generators
$L_x$, $L_y$, $L_z$ of O(3). If we have a state $|\psi \rangle$ which is a 
superposition of the states with $\Lambda = \mbox{${1\oo 2}$}$ and
$\Lambda = -\mbox{${1\oo 2}$}$,
\be
     |\psi \rangle = A |lm,\mbox{${1\oo 2}$} \rangle +
     B |lm,\mbox{$-{1\oo 2}$} \rangle
\lbl{71}
\ee
then an element $S$ of the group SU(2) changes the coefficients $A$, $B$
into new coefficients $A'$, $B'$, so that the new state is
\be
     |\psi' \rangle = S |\psi \rangle = A' |lm,\mbox{${1\oo 2}$} \rangle +
     B' |lm,\mbox{$-{1\oo 2}$} \rangle 
\lbl{72}
\ee

That an extra SU(2) group is present in our representation of
spin ${1\oo 2}$ states is very interesting. It would be challenging to
investigate whether the group SU(2) generated by $T_\alpha$ has
any relation with weak interactions and whether the states
$\chi_{lm}$ and $\theta_{lm}$, $l={1\oo 2}, ~m=\pm {1\oo 2}$ could
represent the weak interaction doublet, with the difference that
they cannot be directly identified with electron $e$ and
neutrino $\nu_e$. Wave functions for the realistic
electron and neutrino would take place in a full relativistic theory.
A step into this direction is provided in next section.

\section{Rigid Particle}

The so called ``rigid particle" which is described by the action containing
second order derivatives (extrinsic curvature) has attracted much attention.
\ci{Rigid}--\ci{PavsicRigidConsist,Kosyakov}. 
Such particle follows in general a worldline which deviates from a
straight line. According to the terminology used in a recent review
\ci{Kosyakov} it exhibits {\it non Galilean motion} which manifests itself
as {\it Zitterbewegung} responsible for particle's spin. Hence, although
the particle is point like it possesses spin.

We are now going to present a revisited review of the rigid particle
with the square of the extrinsic curvature in the action and show that
according to the findings of Sec. 2 the rigid particle can have integer and
half integer spin values.

\subsection{Clasical rigid particle}

\subsubsection{The action and equations of motion}

We shall consider the free rigid particle in Minkowski spacetime with
the metric $g_{\mu \nu} = {\rm diag} (+ - - -)$. The action is 
\ci{PavsicRigid1,PavsicRigid2}
\be
    I = \int \dd \tau \, \gamma^{1/2} (m - \mu H^2) \; , \qquad \gamma \equiv
    {\dot x}^\mu {\dot x}_\mu
\lbl{3.1}
\ee
where $m$ and $ \mu$ are constants, the bare mass and rigidity, respectively;
$\tau$ is an arbitrary monotonically increasing
parameter on the worldline, ${\dot x}^\mu \equiv
\dd x^\mu/\dd \tau, ~ H^2 \equiv g_{\mu \nu} H^\mu H^\nu$, and
\be
    H^\mu \equiv {{\DD^2  x^\mu}\oo {\DD \tau^2}} \equiv {1\oo {\gamma^{1/2}}}
    \, {\dd \oo {\dd \tau}} \left ( {{{\dot x}^\mu}\oo {\gamma^{1/2}}} \right )
    \equiv {{\dd^2 x^\mu }\oo {\dd s^2}} \; , 
    \qquad \dd s = \gamma^{1/2} \, \dd \tau
\lbl{3.2}
\ee
From the action (\ref{3.1}) one can derive, besides the usual pair
of canonically conjugate variables $(x^\mu, p_\mu)$ also the pair
$({\dot x}^\mu, \pi_\mu)$, where $\pi_\mu = - (2 \mu/\gamma^{1/2}) H_\mu$.
The ``internal" space here consists of velocites and the corresponding
conjugate momenta $\pi_\mu$.

A classically equivalent action that was considered by Lindstr\" om 
\ci{LindstromAccel}
is
\be
     I[]x^\mu,y^\mu] = \int \dd \tau \, \gamma^{1/2} \left [ m - \mu \left (
     y^\mu y_\mu - 2 {{{\dot x}^\mu {\dot y}_\mu}\oo \gamma} -
     {{({\dot x}^\mu y_\mu)^2}\oo \gamma} \right ) \right ]
\lbl{3.3}
\ee
The latter action is invariant under reparametrizations of $\tau$ and also
under an extra gauge symmetry discussed by Lindstro\" om \ci{LindstromAccel}:
\be
   y^\mu \rightarrow y^\mu + v(\tau) {\dot x}^\mu
\lbl{3.3a}
\ee
where $v(\tau)$ is an arbitrary function.

Varying the action (\ref{3.3}) with respect to $x^\mu$ and $y^\mu$ we obtain
($\gamma \equiv {\dot x}^2$):
\bear
    &&\delta x^\mu \, : \quad {\dot p}_\mu = 0 \lbl{3.4a} \\
    &&\delta y^\mu \, : \quad {\dot P}_\mu = 2 \mu \gamma^{1/2} \left ( y_\mu
    - {1\oo \gamma} \, ({\dot x}^\nu y_\nu) \, {\dot x}^\mu \right ) \lbl{3.4b}
\ear
where
\be
  p_\mu = {{\p L}\oo {\p {\dot x}^\mu}} =   
   {{m {\dot x}_\mu}\oo {\gamma^{1/2}}} - {\mu\oo {2 \gamma^{3/2}}}
   \left [ \gamma y^2 {\dot x}_\mu + 2 {\dot x}^\nu {\dot y}_\nu \, {\dot x}_\mu
   + 2 \gamma {\dot y}_\mu +  ({\dot x}^\nu y_\nu )^2 {\dot x}_\mu
   - 2 \gamma {\dot x}^\nu y_\nu \, y_\mu \right ]
   \lbl{3.5a}
\ee
\be   
   P_\mu = {{\p L}\oo {\p {\dot y}^\mu}} = - {{2 \mu \, {\dot x}_\mu} \oo
   \gamma^{1/2}} \hs{10.4cm}
\lbl{3.5b}
\ee
From the equation of motion (\ref{3.4b}) we find the relation
\be
      y^\mu = {{{\ddot x}^\mu} \oo {{\dot x}^2}}
\lbl{3.4c}
\ee
We see that $y^\mu$ is proportional to the acceleration ${\ddot x}^\mu$,
whilst $P^\mu$ is proportional to the velocity ${\dot x}^\mu$.

The pairs of canonically conjugate variables are 
\be
      (x^\mu , p_\mu) \qquad {\rm and} \qquad (y^\mu,P_\mu)
\lbl{3.5}
\ee

The generators of infinitesimal translations $x^\mu \rightarrow x^\mu +
\epsilon^\mu$ and rotations $x^\mu \rightarrow x^\mu + {\epsilon^\mu}_\nu
x^\nu$, $y^\mu \rightarrow y^\mu +{\epsilon^\mu}_\nu y^\nu$, $\epsilon_{\mu \nu}
= - \epsilon_{\nu \mu}$ are $p_\mu$ and $J_{\mu \nu} = M_{\mu \nu} +
S_{\mu \nu}$, respectively, where
\be
       M_{\mu \nu} = x_\mu p_\nu - x_\nu p_\mu
\lbl{3.6}
\ee
\be
     S_{\mu \nu} = y_\mu P_\nu - y_\nu P_\mu
\lbl{3.7}
\ee
are orbital angular momentum and spin tensor. Occurrence of spin in our
dynamical system results from the curvature term in the action (\ref{3.1}),
or equivalentrly, from the terms with $y^\mu$ in the action (\ref{3.3}).
A result is that the particle does not follow a straight world line, but
performs a Zitterbewegung.

The canonical momenta $p_\mu$ and $P_\mu$ satisfy the following
two constraints \ci{LindstromAccel}:
\bear
        && \phi_2 \equiv p_\mu P^\mu - {{\mu m}\oo 2} - S_{\mu \nu} S^{\mu \nu}
    = 0 \lbl{3.8}\\
    &&\phi_1 \equiv P_\mu P^\mu - \mu^2 = 0 \lbl{3.9} 
\ear
which are due to the invariance of the action (\ref{3.3}) under
reparametrisations of $\tau$, and under the transformation (\ref{3.3a}).
The constraints (\ref{3.8}),(\ref{3.9}) are first class, because
their Poisson bracket is strongly zero:
\be
     \lbrace \phi_1,\phi_2 \rbrace = 0
\lbl{3.10}
\ee

The Hamiltonian is a linear combination of constraints:
\be
     H = v_1 \phi_1 + v_2 \phi_2
\lbl{3.11}
\ee
and it generates the $\tau$-evolution of an arbitrary quantity
$A(x^\mu,p_\mu,y^\mu,P_\mu)$ of the canonically conjugate variables
$x^\mu,p_\mu,y^\mu,P_\mu$. So we obtain that the total angular momentum       
$J_{\mu \nu}$ is a constant of motion:
\be
    {\dot J}_{\mu \nu} = \lbrace J_{\mu \nu},H \rbrace = 0
\lbl{3.12}
\ee
where the dot denotes the derivative with respect to $\tau$.

Another quantities which are also conserved are the 
{\it Pauli-Lubanski pseudo vector}
\be
     S^\mu = {1\oo \sqrt{p^2}} \, \epsilon^{\mu \nu \alpha \beta}
     p_\nu J_{\alpha \beta} = 
     {1\oo \sqrt{p^2}} \, \epsilon^{\mu \nu \alpha \beta} p_\nu S_{\alpha \beta}
\lbl{3.12a}
\ee
and the {\it momentum} $p_{\mu}$ (defined in (\ref{3.5a})). Thus
\be
    {\dot p}_\mu =\lbrace p_\mu,H \rbrace = 0
\lbl{3.12b}
\ee
\be
    {\dot S}^\mu = \lbrace S^\mu,H \rbrace = 0
\lbl{3.14}
\ee

But the momentum $P_\mu$, conjugate to $y^\mu$, is {\it not} conserved:
\be
     {\dot P}_\mu = \lbrace P_\mu,H \rbrace \neq 0
\lbl{3.15}
\ee
where the right hand side of the latter equation is given in eq.(\ref{3.4b}).

\subsection{Quantization}

The system can be quantized by replacing the canonically conjugate
pairs of variables $(x^\mu,p_\mu)$ and $(y^\mu,P_\mu)$ by operators
satisfying the following commutation relations\footnote{We use the units
in which $\hbar = c = 1$.}
\be
     [x^\mu,p_\nu] = i {\delta^\mu}_\nu \; , \qquad [y^\mu,P_\nu] =
         i {\delta^\mu}_\nu
\lbl{3.16}
\ee
The constraints (\ref{3.8}),(\ref{3.9}) become the conditions a physical
state has to satisfy:
\bear
      &&\phi_1 \psi \equiv (P^\mu P_\mu - \mu^2) \, \psi = 0 \lbl{3.17} \\
      &&\phi_2 \psi \equiv \left ( p_\mu P^\mu - {\mu m\oo 2} - 
      S_{\mu \nu} S^{\mu \nu} \right )
      \psi = 0 \lbl{3.18}
\ear
We find $[\phi_1,\phi_2] = 0$ which assures that the conditions
(\ref{3.17}),(\ref{3.18}) are consistent.

The momentum $p_\mu$ and the Pauli-Lubanski operator $S^\mu$ commute with
the operators $\phi_1$ and $\phi_2$:
\be
      [p_\mu,\phi_1] = 0 \; , \qquad [p_\mu, \phi_2] = 0
\lbl{3.19}
\ee
\be
       [S^\mu,\phi_1] = 0 \; , \qquad [S^\mu,\phi_2] = 0
\lbl{3.20}
\ee

The set of mutually commuting operators is
$\lbrace p_\mu,~S^\mu,~\phi_1,~\phi_2 \rbrace$.
They can thus have simultaneous eigenstates
and eigenvalues. The physical states can be classified by the eigenvalues
of the mass squared operator $p^\mu p_\mu$ and spin $S^\mu S_\mu$.
Eigenvalues of the spin operator $S^\mu S_\mu$ are $s(s+1)$. Choosing a
representation in which $x^\mu$ and $y^\mu$ are diagonal, the corresponding
momenta and spin are differential operators 
\be
  p_\mu = - i \p/\p x^\mu \quad {\rm and} \quad  P_\mu = - i \p/\p y^\mu
\lbl{3.20a}
\ee
\be
    S_{\mu \nu} = P_\mu y_\nu - P_\nu y_\mu
\lbl{3.20b}
\ee       
Assuming that $\psi$ are eigenfunctions of the momentum $p^\mu$ and
that a reference frame exists in which $p^\mu = (p^0,0,0,0)$, we find
that the equations
\be
     S^\mu S_\mu \psi = s(s+1) \psi \lbl{3.21} \ee
\be     
     S_z \psi = s_z \psi       \lbl{3.22} \ee
become differential equations equivalent to the equations (\ref{2}),(\ref{3}).
     
Eq. (\ref{3.17}) becomes the differential equation and can be reduced
to a form which is mathematically equivalent to the static Schr\" odinger
equation and which in spherical coordinates leads to the equation for the
eigenfunctions of the angular momentum operator.

The formalism describing the rigid particle thus becomes equivalent
to the formalism of Sec.\,2, where we considered the Schr\" odinger basis
for spinor representation of the rotation group. In rigid particle
we have a concrete physical realization the Schr\" odinger basis for
spin which, as we have shown in Sec.\,2, allows for integer and
half-integer spin values.

Although the spin angular momentum $S_{\mu \nu}$ formally looks like
the orbital momentum operator, there is a big difference.

In the case of a free {\it point particle}, its momentum $p_\mu$ is
a constant of motion, and so are its orbital angular momentum squared and $L_z$.
Therefore, a state of a free particle can be expanded either in
terms of the momentum eigenfucntions or equivalently, in terms of the
orbital angular momentum eigenfunctions. Momentum eigenfunctions form a complete
set of states, and they are single valued. Therefore, when using the orbital
angular momentum
eigenfucntions one has to take into account only {\it single valued}
functions. The orbital angular momentum of a point particle
has thus integer values only. Such argument was provided in Dirac's book
on quantum mechanics \ci{DiracBook}.

In the case of {\it rigid particle} the role that $p_\mu,~x^\mu$ had in
Sec.2 is assumed
by $P_\mu,~y^\mu$. But $P_\mu$, unlike $p_\mu$, is {\it not} a constant
of motion\footnote{If we switch off the rigidity by setting the rigidity
constant $\mu$ equal to zero,
then, according to eq.\,(\ref{3.5b}), $P_\mu = 0$, which is a trivial constant.
For non vanishing rigidity constant $\mu$ we have that in general
$P_\mu$ differs from
zero; and if it differs form zero, then automatically it cannot be a
constant of motion. In this respect rigid particle is drastically different
from the ordinary particle. If an ordinary particle moves in the presence
of a spherically symmetric potential, then, of course, its linear momentum
$p_\mu$ is not a constant of motion, while its angular momentum is constant.
But if one switches off the potential, $p_\mu$ becomes constant that
can differ from zero.}.
A state of the rigid particle cannot be described as a
superposition of the eigenstates of $P_\mu$. However, it can be
described as a superposition of the eigenstates of $S^\mu S_\mu$ and
$S_z$, which are constants of motion, and which, as shown
in sec.\,2, can have eigenvalues either for integer or half-integer $l$.
In other words,
the linear momentum $P_\mu$ (which is conjugate to the acceleration), does
not commute with the Hamiltonian operator $H=v_1 \phi_1 + v_2 \phi_2$,
hence it is not a constant of motion, and therefore the eigenfunctions of
$P_\mu$, which are single-valued only, cannot serve
for a description of the rigid particle. On the other hand, the mutually
commuting operators, $S^\mu S_\mu,~ S_z$ and H do have the simultaneous
eigenfunctions. The latter eigenfunctions do provide a description
of the rigid particle, and they may be single or double valued.

\section{Conclusion}

We have clarified a long standing problem concerning the admissibility
of double valued spherical harmonics in providing a spinor representation
of the three dimensional rotation group. 
The usual arguments against the inadmissibility of such functions,
concerning hermiticity, orthogonality, behaviour under rotations, etc.,
are  all related to the unsuitable choice of functions representing the
states with positive and negative values of the quantum number $m$,
and to an inappropriate definition of inner product.
By the correct choice of functions such problems do not occur, provided
that we modify the inner product as well. We have considered two
different definitions of inner product. By using Definition I the
spherical harmonics with half-integer spin values do form
a reducible representation of the 3-dimensional rotation group.
But the latter representation is not fully reducible, because the
physical space ${\cal S}_l$ with $m$-values in the range between $l$
and $-l$ is not invariant under rotations. But because the states projected
onto ${\cal S}_l$ transform in the correct way with their norms 
preserved, such representation makes sense.
This is explicitly illustrated on an example for
$m = +{1\oo 2},-{1\oo 2}$.
With Definition II of inner product
spherical harmonics form a fully reducible representation of
rotation group even for half-integer spin values.

Double valued spherical harmonics are admissible, 
if they do not refer to the ordinary configuration space in
which the usual  quantum mechanical orbital angular momentum is defined,
but if they refer to an internal space in which a spin angular momentum
is defined. An example of such an internal space is the space of
velocities, or, equivalently, the space of accelerations, associated
with the so called {\it rigid particle}
whose action contains the square of the
extrinsic curvature of a particle's
world line. If one considers the
action (\ref{3.1}), then one has the space of velocities. But in several
respects it is more convenient to consider an alternative, although classically
equivalent action (\ref{3.3}), in which case the internal space consists
of accelerations.

\centerline{\bf Acknowledgement}

This work has been supported by the Ministry of
High Education, Science and Technology of Slovenia, grant No. PO-0517.

\end{document}